\documentclass[11pt,a4paper]{article}
\pdfoutput=1
\usepackage{jcappub}
\usepackage{aasmacros}

\ifx\pdfoutput\undefined
\usepackage[dvips,bookmarks=false]{hyperref}	
\else
\usepackage{hyperref}	
\fi
\hypersetup{colorlinks,bookmarksopen,bookmarksnumbered,citecolor=blue,
linkcolor=black,pdfstartview=FitH,urlcolor=blue}

\usepackage[title]{appendix}
\usepackage{graphicx}
\usepackage{amsmath}
\usepackage{amssymb}
\usepackage{rotating} 
\usepackage{xspace}

\usepackage{placeins}
\usepackage{rotating}

\newcommand{\be}{\begin{equation}}
\newcommand{\ee}{\end{equation}}
\newcommand{\beq}{\begin{equation}}
\newcommand{\eeq}{\end{equation}}

\newcommand{\vect}[1]{\boldsymbol{\rm #1}}

\newcommand{\Msun}{\rm{M}_{\odot}}

\makeatletter
\renewcommand{\fnum@table}{\textbf{\tablename~\thetable}}
\renewcommand{\fnum@figure}{\textbf{\figurename~\thefigure}}
\makeatother

\title{Probing the nature of dark matter particles with stellar streams}
\author[a,b]{Nilanjan Banik,}
\author[a]{Gianfranco Bertone,}
\author[c,d]{Jo Bovy}
\author[a,e]{and Nassim Bozorgnia}

\affiliation[a]{GRAPPA Institute, Institute for Theoretical Physics Amsterdam\\ 
and Delta Institute for Theoretical Physics, University of Amsterdam, \\
Science Park 904, 1098 XH Amsterdam, The Netherlands
} 

\affiliation[b]{Lorentz Institute, Leiden University, Niels Bohrweg 2,\\
Leiden, NL-2333 CA, The Netherlands} 

\affiliation[c]{Department of Astronomy and Astrophysics, University of Toronto,\\ 
50 St. George Street, Toronto, ON, M5S 3H4, Canada
} 

\affiliation[d]{Alfred P. Sloan Fellow}

\affiliation[e]{Institute for Particle Physics Phenomenology, Department of Physics,\\ 
Durham University, Durham, DH1 3LE, United Kingdom
}

\emailAdd{banik@lorentz.leidenuniv.nl}

\abstract{A key prediction of the standard cosmological model -- which relies on the assumption that dark matter is cold, i.e. non-relativistic at the epoch of structure formation -- is the existence of a large number of dark matter substructures on sub-galactic scales. This assumption can be tested by studying the perturbations induced by dark matter substructures on cold stellar streams. Here, we study the prospects for discriminating cold from warm dark matter by generating mock data for upcoming astronomical surveys such as the Large Synoptic Survey Telescope (LSST), and reconstructing the properties of the dark matter particle from the perturbations induced on the stellar density profile of a stream. We discuss the statistical and systematic uncertainties, and show that the method should allow to set stringent constraints on the mass of thermal dark matter relics, and possibly to yield an actual measurement of the dark matter particle mass if it is in the $\mathcal{O}(1)$ keV range.
}

\arxivnumber{IPPP/18/24}

\begin{document}
\maketitle
\section{Introduction}
\label{sec:introduction}

Understanding the nature of dark matter is one of the most pressing problems in cosmology and particle physics~\cite{Bertone:2010zza,Jungman:1995df,Bergstrom00,Bertone05}. A key prediction of the standard cosmological model, based on the assumption that dark matter is cold, i.e. non-relativistic at the epoch of structure formation, is that a large number of dark matter substructures should exist in Milky Way-like galaxies \cite{Diemand2008,Springel2008}. 
Detecting these subhalos will not only be a strong indicator for the existence of dark matter but also will give valuable information about its particle nature. Depending on their couplings and production mechanism, in fact, dark matter particles can achieve non-negligible velocity dispersions -- thus act as warm dark matter (WDM) -- leading to a suppression of the primordial density fluctuations on small scales, and thus a cutoff at small halo masses, as already realised long ago \cite{1982ApJ...258..415P}. 

We discuss below the case of a thermal WDM relic, for which the cutoff in the power spectrum depends only on the WDM particle mass, but the discussion can be generalised to sterile neutrinos \cite{Dodelson1994,Shi:1998km,Abazajian:2001nj,Asaka2005,Boyarsky2009,Adhikari:2016bei} by taking into account the lepton asymmetry parameter (see next section). 
The subhalo mass function of WDM and cold dark matter (CDM) differs for masses smaller than those of dwarf galaxies. On these scales,  such substructures would be completely dark matter dominated, and contain too few stars to be observed directly. A number of strategies have been proposed to indirectly probe low mass dark matter halos, including analyses of the impact on Ly$\alpha$ forest observations \cite{Narayanan:2000tp,Viel2005,Boyarsky:2008xj,Viel2013,Baur:2015jsy,Garzilli:2015iwa,Irsic:2017ixq} and perturbations of strong gravitational lenses \cite{Dalal2002,Vegetti:2008eg,Li:2015xpc,Penarrubia:2017nzw,Asadi:2017ddk,Mao:2017auo,Daylan:2017kfh,Minor:2016jou,Despali:2016meh}. 

Here, we study the prospects for identifying the nature of dark matter particles by studying the perturbations induced by sub-dwarf galaxy clumps on cold stellar streams. A stellar stream, created as a result of tidal disruption of a globular cluster or dwarf galaxy by the Milky Way potential, has more or less uniform stellar density along its length. A flyby subhalo gravitationally perturbs the stars in the stream resulting in a region of low stellar density or gap whose size increases with time. There has been a lot of work recently \cite{Yoon2011,Carlberg2012,Carlberg2013,Erkal2015,Erkal2015a,Sanders2016} focusing on the study of gaps in stellar streams as a result of subhalo encounters. Specifically, ref.~\cite{Erkal2015a} showed that with accurate measurements of the density and phase-space structure of stellar streams, which is possible with the near future galaxy surveys like LSST and the Gaia mission currently taking data, we will be able to measure subhalo impacts with mass as low as $10^7 ~\Msun$. In ref.~\cite{Bovy2016a}, a novel framework was presented for inferring properties of the impacting subhalos by analyzing the power spectrum of the fluctuations in the density and mean track of the perturbed stream. This method was applied on Pal 5 density data and was shown to be sensitive to subhalos with mass as low as $10^{6.5} ~ \Msun$. The authors also predicted that with better data this method will be sensitive to even $10^5 ~\Msun$ subhalos. In order to discriminate between WDM and CDM, we will in particular exploit the difference in the density power spectrum arising from the different subhalo populations. 

This paper is structured as follows. In section \ref{sec:WDM}, we describe the WDM model that we used to simulate the stream-subhalo encounters in the WDM scenario. In section \ref{sec:simstream}, we discuss our method of generating fast stream density simulations of the GD-1 stream and how we model subhalo impacts on it. In section \ref{sec:results}, we present our results of the stream density perturbations and their power spectra, and apply the Approximate Bayesian Computation (ABC) technique to constrain the mass of the dark matter particle from the stream power spectra. We show that for typical cases we are able to distinguish between WDM and CDM models. In section \ref{sec:outlier}, we demonstrate the risks of using only one stream to constrain the particle mass of dark matter. With these examples we make a case for the need of applying our method on multiple streams to make robust predictions on the mass of the dark matter particle. Finally we conclude in section \ref{sec:conclusions}. In appendix \ref{sec:scaleradius} we discuss how the scale radius of subhalos in the WDM scenario differ from the CDM scenario and why this variation has little effect on our method.        

\section{Modeling warm dark matter}
\label{sec:WDM}

We discuss below a thermal WDM relic, for which the cutoff in the power spectrum depends only on the WDM particle mass, but the discussion can be generalised to sterile neutrinos by taking into account the lepton asymmetry parameter\footnote{The lepton asymmetry is defined as $L_6 \equiv 10^6 (n_{\nu_e} - n_{\bar{\nu_e}})/s$, where $n_{\nu_e}$  and $n_{\bar{\nu_e}}$ are the number densities in electron neutrinos and anti-neutrinos respectively, and $s$ is the entropy density of the Universe. For very high and low values of lepton asymmetry the power spectrum of sterile neutrinos can be well approximated by that of a thermal WDM.}. For example, the power spectrum of a 3.3 keV thermal WDM matches to a high degree with that of a 7 keV sterile neutrino with lepton asymmetry parameter equal to 8.66 (see figure 1 in ref.~\cite{Bose2015}), with both power spectra deviating from the CDM case at around $\log(k) \gtrsim 1.0 ~h~\rm{Mpc}^{-1}$, where $k$ is the comoving wave number and $h=H_0/100$~km$/$s$/$Mpc is the dimensionless Hubble parameter. The 7 keV sterile neutrino is especially interesting since its decay products have been attributed to be the source of the 3.5 keV X-ray line \cite{Bulbul2014,Boyarsky2014} detected in stacked spectrum of clusters as well as in the spectra of Andromeda and Perseus cluster. Notice that the case of the 7 keV sterile neutrino with lepton asymmetry of 8.66 corresponds to the coldest in the 7 keV  sterile neutrino family, so any constraints on it can be extended to all the other members of the family. 

Various studies have set a  lower bound on the mass of thermal WDM based on the observed clumpiness in the Ly-$\alpha$ forest. Ref.~\cite{Viel2013} found this lower bound to be 3.3 keV ($2\sigma$), while ref.~\cite{Baur:2015jsy} found it to be 4.09 keV (95\% CL) from Ly-$\alpha$ data alone, and 2.96 keV (95\% CL) when combining Ly-$\alpha$ data with CMB data from Planck 2016. Very recently, ref.~\cite{Yeche:2017upn} found this lower limit to be 4.17 keV (95\% CL) using two different Ly-$\alpha$ measurements, and 4.65 keV (95\% CL) when adding a third set of data. Notice that the constraints on the thermal WDM mass depend on the thermal history assumed for the intergalactic medium~\cite{Garzilli:2015iwa}.

In this work we model WDM as a thermal relic and emphasize on the case of 3.3 keV particle mass. We follow the same framework that was adopted in WDM simulation works \cite{Lovell2013,Bose2015,Bose2016}. The WDM and CDM power spectra are related by a transfer function $T(k)$:
\begin{equation}
P_{\rm{WDM}} (k) = T^{2}(k)P_{\rm{CDM}}(k),
\end{equation}
where $T(k)$ is approximated by the fitting formula \cite{Bode2001}:
\begin{equation}
T(k) = (1+(\alpha k)^{2\nu})^{-5/\nu},
\end{equation} 
with $\nu$ and $\alpha$ constants. Ref.~\cite{Viel2005} found that for $k < 5~ h~\rm{Mpc}^{-1}$, the best fit of the transfer function is obtained with $\nu = 1.12$.  $\alpha$ is the cutoff scale as a result of free streaming in the WDM power spectrum relative to CDM, which following ref.~\cite{Viel2005}, is given by
\begin{equation}
\alpha = 0.047\left(\frac{m_{\rm{WDM}}}{\rm{keV}}\right)^{-1.11}\left(\frac{\Omega_{\rm{WDM}}}{0.2589}\right)^{0.11}\left(\frac{h}{0.6774}\right)^{1.22}h^{-1}~\rm{Mpc}.
\end{equation}
Here $m_{\rm{WDM}}$ is the WDM particle mass and $\Omega_{\rm{WDM}}$ is the WDM contribution to the density parameter.

The length scale at which the transfer function drops by a factor of 2 is the \textit{half mode} wavelength, $\lambda_{\rm{hm}}$, and the mean mass enclosed within a sphere of diameter equal to $\lambda_{\rm{hm}}$ is the \textit{half mode} mass, $M_{\rm{hm}}$. The \textit{half mode} mass quantifies the threshold mass below which the WDM subhalo mass function is strongly suppressed. For example, for a 3.3 keV thermal WDM, $M_{\rm{hm}} \sim 2 \times 10^{8} h^{-1} \rm{M}_{\odot}$ \cite{Colin2008,Angulo2013,Viel2013}.

Based on this WDM model, ref.~\cite{Schneider2012} performed a series of
high-resolution N-body simulations and obtained a functional fit for the differential mass function of WDM relative to CDM as
\begin{equation}
\left(\frac{dn}{dM}\right)_{\rm{WDM}} = \left(1+ \frac{M_{\rm{hm}}}{M}\right)^{-\beta} \left(\frac{dn}{dM}\right)_{\rm{CDM}},
\end{equation} 
where $M$ is the subhalo mass and $\beta$ is a free parameter found to be equal to 1.16. Ref.~\cite{Lovell2013} studied the abundance and structure of WDM subhalos within a Milky Way-like host halo using high resolution N-body simulations based on the Aquarius Project \cite{Springel2008}. They found that the above functional fit is improved by introducing another parameter $\gamma$, such that
\begin{equation}
\left(\frac{dn}{dM}\right)_{\rm{WDM}} =  \left(1+ \gamma\frac{M_{\rm{hm}}}{M}\right)^{-\beta} \left(\frac{dn}{dM}\right)_{\rm{CDM}},
\label{eq:wdmcdm}
\end{equation}
with $\gamma = 2.7$ and $\beta = 0.99$. 

 Since the number of subhalo encounters experienced by a particular stellar stream depends on the number of subhalos within the average radius of the stream from the galactic center, we need a model for the radial dependence of the number density of WDM subhalos. In ref.~\cite{Springel2008}, it was shown that for a Milky Way like host galaxy, the radial distribution of CDM subhalos in the mass range $[10^5 - 10^9]~ \Msun $  is described by an Einasto profile. Furthermore, the CDM subhalo mass function for those halos were shown to be well described by $dn/dM \propto M ^{-1.9}$. These results were combined in ref.~\cite{Erkal2016} to estimate the normalized subhalo profile at a given mass and at a certain galactrocentric radius:
\begin{equation}
\left(\frac{dn}{dM}\right)_{\rm{CDM}} = c_{0}\left(\frac{M}{m_{0}}\right)^{-1.9}\exp\left\{ - \frac{2}{\alpha}\left [\left(\frac{r}{r_{-2}}\right)^{\alpha} - 1 \right]\right\}
\label{eq:dndMc}
\end{equation}
with $c_{0}=2.02 \times 10^{-13}~\Msun^{-1}$~kpc$^{-3}$, $m_{0}= 2.52\times 10^{7}~\Msun$, $\alpha = 0.678$ and $r_{-2} = 162.4$~kpc. 

Ref.~\cite{Lovell2013} found that the radial distributions of WDM subhalos of different particle masses are very similar to one another and to the CDM case. We can therefore adopt the Einasto profile for the radial distribution of WDM subhalos. We combine eqs.~\eqref{eq:wdmcdm} and \eqref{eq:dndMc} to obtain an expression for the WDM subhalo profile,
\begin{equation}
\left(\frac{dn}{dM}\right)_{\rm{WDM}} = c_{0}\left(\frac{M}{m_{0}}\right)^{-1.9}\exp\left \{ - \frac{2}{\alpha}\left[\left(\frac{r}{r_{-2}}\right)^{\alpha} - 1 \right]\right \}\left(1+ \gamma\frac{M_{\rm{hm}}}{M}\right)^{-\beta}.
\label{eq:dndMw}
\end{equation} 

Notice that we have ignored the evolution of the subhalo number density over the age of the stream, since we do not expect it to cause any significant change in the results. This is because gaps fill up over time due to the internal velocity dispersion in the stream and therefore very old gaps will not be visible today.

Baryonic effects have been shown to tidally disrupt dark matter subhalos thereby reducing their number. Ref.~\cite{Sawala2016} used APOSTLE simulations in the $\Lambda$CDM framework to estimate that the substructure abundance in the mass range $[10^{6.5} - 10^{8.5}]~\Msun$ inside a Milky Way mass halo is greatly affected over a lookback time of up to 5 Gyr. Within the radius of the GD-1, they predicted a reduction of dark matter substructures by $\sim 45 - 50 \%$. Taking this into account, we reduced the number of substructures in each mass decade within the subhalo mass range $[10^6 - 10^9]~\Msun $ by 47\%. For WDM subhalos this factor may be even greater as they are more easily tidally disrupted owing to their lower concentration at the time of their infall. However, we have ignored this in the present work. 

\begin{figure}[t]
\centering
\includegraphics[scale=0.65]{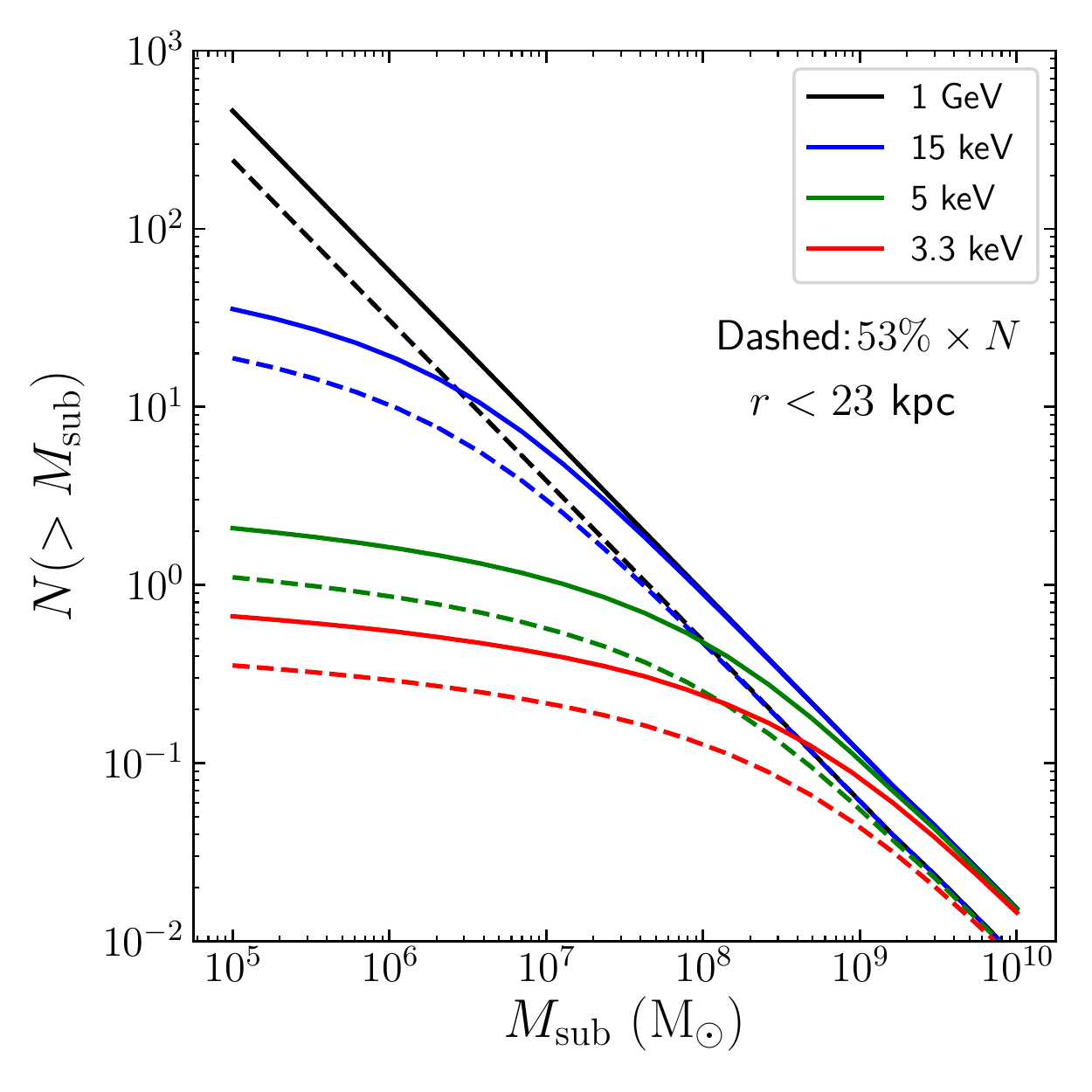}
\caption{Cumulative number of subhalos within a Galactocentric radius of 23 kpc of a Milky Way sized host halo for different WDM particle mass scenarios. Dashed lines indicate the cumulative numbers taking into account the 47\% reduction of the number of subhalos due to baryonic effects.}

\label{fig:Nsub}
\end{figure}

Figure \ref{fig:Nsub} shows the cumulative number of subhalos for different WDM particle masses obtained by integrating eq.~\eqref{eq:dndMw} up to a subhalo mass of $10^{12}~ \Msun$, shown by the solid lines. The dashed lines represent the cumulative number of subhalos taking the 47\% reduction into account.

\section{Modeling the GD-1 stream and subhalo impacts}
\label{sec:simstream}

\subsection{Generating a smooth stream}
\label{subsec:smoothstream}
In order to analyze the gap power spectrum of a stellar stream due to impacts from thermal WDM subhalos, we need a mock stream that had minimal gap inducing perturbations from baryonic effects. In refs.~\cite{Erkal2017,Pearson2017}, it was shown that the Milky Way bar can induce gaps in stellar streams such as Pal 5 \cite{Odenkirchen2001}, which is in prograde motion with respect to the pattern speed of the bar. Such gaps are not induced in streams like the GD-1 \cite{Grillmair2006} which is in retrograde motion with respect to the bar \cite{Koposov2010, Bovy2016,Erkal2017}. Furthermore, ref.~\cite{Amorisco2016} studied the effects of giant molecular clouds on the GD-1 stream and found that because of the giant molecular clouds'  larger pericentre and steeper mass function, and also due to GD-1's retrograde motion there is no strong density perturbations induced on the GD-1 stream. Taking all these points into consideration, the density gaps found in GD-1 stream are expected to be solely due to dark matter subhalo impacts. Therefore, we generate a mock GD-1 stream  for our present analysis.   

To generate a mock GD-1 stream and simulate the impacts due to dark matter subhalos, we make use of the simple model of stream evolution and subhalo impacts in the space of orbital frequency $\boldsymbol{\Omega}$, and orbital angle $\boldsymbol{\theta}$, that was developed in ref.~\citep{Bovy2014} and is included in the \texttt{galpy} code \cite{Bovy2015}. For a detailed step-by-step explanation of this entire method see refs.~\cite{Bovy2014,Bovy2016a}. We briefly summarize the method here. Given a model host potential, current progenitor phase-space information, velocity-dispersion parameter $\sigma_{v}$ of the progenitor, and a disruption time $t_d$ at which disruption started, a leading
or trailing tail model of the stream is generated in ($\boldsymbol{\Omega},\boldsymbol{\theta}$) space. In this work we use the well-tested three component Milky Way potential $\texttt{MWPotential2014}$ from ref.~\cite{Bovy2015} as the host potential. The GD-1 progenitor's phase space coordinate was taken from ref.~\cite{Bovy2016} : $(\phi_{1}, \phi_{2}, D, \mu_{\phi_{1}},\mu_{\phi_{2}},V_{\rm{los}}) = (0^{\circ},-0^{\circ}.82 \pm 0^{\circ}.08, 10.1 \pm 0.2 ~ \rm{kpc}, -8.5 \pm 0.3 ~\rm{mas ~yr^{-1}}, -2.15 \pm 0.10 ~ \rm{mas ~ yr^{-1}}, -257 \pm 5 ~\rm{km ~s^{-1}}  )$, where $\phi_{1}$ and $\phi_{2}$ are custom sky coordinates as used in ref.~\cite{Koposov2010}, $D$ is the distance to the progenitor, $\mu_{\phi_{1}}$ and $\mu_{\phi_{2}}$ are the proper motion along $\phi_{1}$ and $\phi_{2}$, and $V_{\rm{los}}$ is the line of sight velocity. It should be noted that the progenitor's location $\phi_{1}$ is set to $0^{\circ}$, $\sigma_{v} = 0.365$ km s$^{-1}$, and we model the whole GD-1 stream as a leading arm. The true extent of the GD1 stream is limited by the edge of current surveys as well as by the galactic disk. Hence, constraining the disruption time $t_d$ is difficult with the present data. We therefore have assumed that the disruption occurred 9 Gyr ago which made our stream very old and long.  

From the progenitor orbit, an approximate Gaussian action ($\textbf{J}$) distribution for the tidally stripped stars is constructed \cite{Eyre2011}, which is then transformed into frequency space using the Hessian matrix evaluated at the progenitor's actions $(\partial \boldsymbol{\Omega}/\partial \textbf{J})_{\mathbf{J_{P}}}$. The resulting variance tensor in frequency space has the principal eigenvector along the direction in frequency space in which the stream spreads, $\boldsymbol{\Omega}_{\parallel}$. The eigenvalues of the eigenvectors perpendicular to $\boldsymbol{\Omega}_{\parallel}$ are less than the largest eigenvalue by a factor of 30 or more \cite{Sanders2013a}. Tidally disrupted stars are generated with a frequency distribution that is modeled as a Gaussian. This Gaussian has a mean equal to the frequency offset from the progenitor and its variance tensor is obtained by transforming the Gaussian action distribution to frequency space and multiplying by the magnitude of the parallel frequency, which is done for the purpose of simplifying analytic calculations. The dispersion of the stellar debris in frequency space is scaled by the velocity dispersion, $\sigma_{v}$, and the relative eigenvalues. Once stripped, the gravitational effects on the stellar debris from its progenitor is neglected and they are evolved solely in their host potential. Their future locations are computed based on their linear evolution in angle space with their frequency remaining constant. For a given disruption time $t_d$, stellar debris is generated with a distribution of constant stripping time with a maximum time $t_d$. Following refs.~\cite{Bovy2014,Bovy2016}, we can transform the stellar debris from ($\boldsymbol{\Omega},\boldsymbol{\theta}$) space to configuration space using linearized transformation near the track of the stream. 

\begin{figure}
\centering
\includegraphics[scale=0.6]{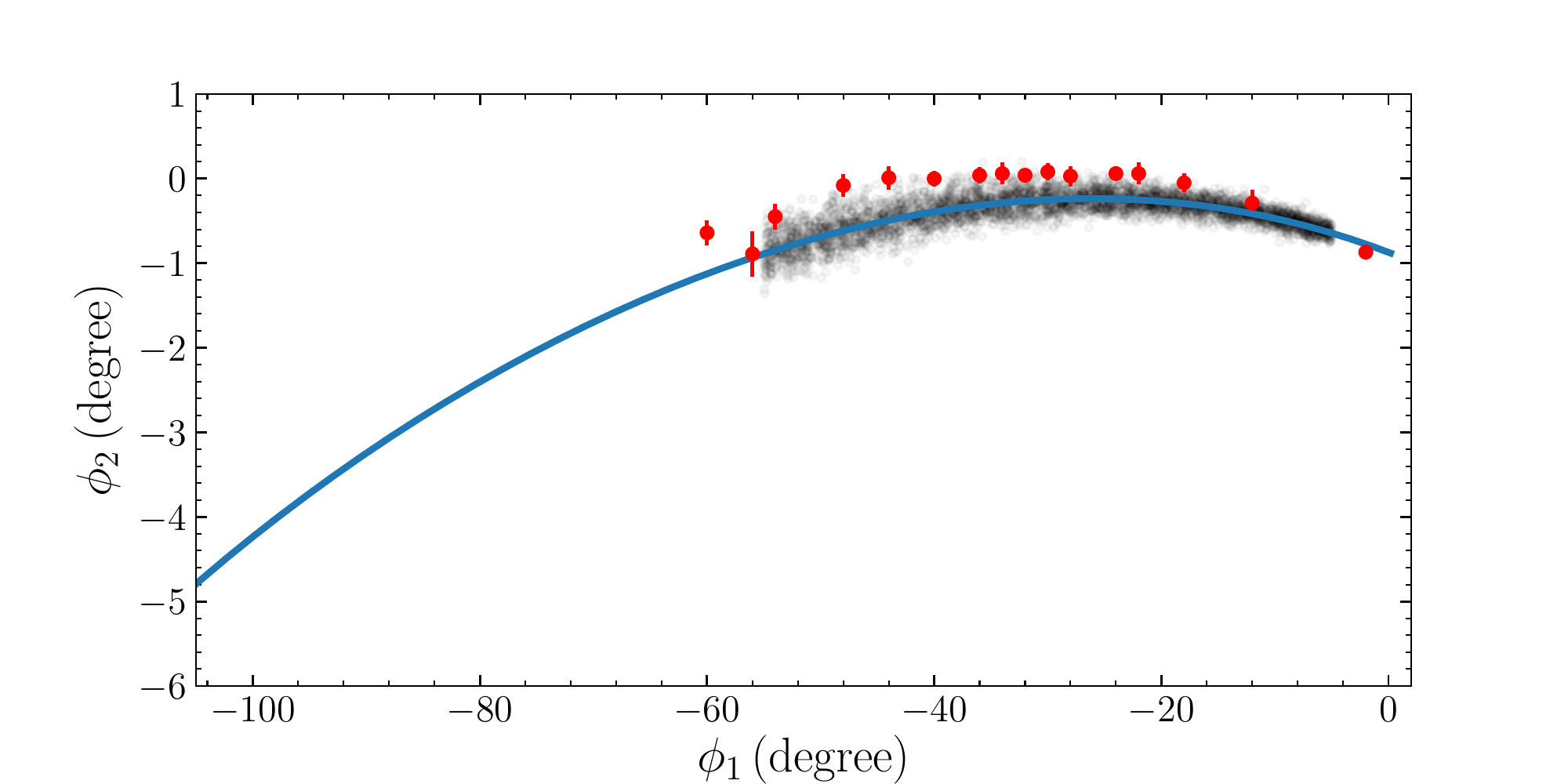}
\caption{GD-1 stream model generated using the framework developed in ref.~\cite{Bovy2014} and using \texttt{galpy}'s \texttt{MWPotential2014} for a stream age of 9 Gyr, velocity-dispersion parameter $\sigma_{v} = 0.365$ km$s^{-1}$ and using the phase-space coordinates from ref.~\cite{Bovy2016}. The blue line is the mean stream track of the full stream in angular coordinates. The gray points show a sampling of mock stream data from the model after applying the angular cut $-55^{\circ} < \phi_{1} < -5^{\circ}$ in order to match the observed stream. The red points are the stream data positions from ref.~\cite{Koposov2010}.}
\label{phi12}
\end{figure}

Based on this approach, we generate a smooth GD-1 stream whose properties, namely path of the smooth stream in Galactic latitude and longitude, parallel angle variation $\Delta\theta_{\parallel}$ with respect to the Galactic longitude, and density variation are similar to those shown in figure 1 of ref.~\cite{Bovy2016a}. The perigalacticon of the stream is $\sim 14$ kpc, the apogalacticon is $\sim 30$ kpc, and the average Galactocentric radius is $\sim 23$ kpc. As shown in figure \ref{phi12}, in the custom sky coordinates $(\phi_{1},\phi_{2})$, the full stream is aligned along $\phi_{1}$ and stretches over $\sim 100^{\circ}$, while along $\phi_{2}$ the stream only extends over $\sim 5^{\circ}$. The gray points are samples drawn from the smooth stream model after applying the angular cut $-55^{\circ} < \phi_{1} < -5^{\circ}$ to match the observed data from ref.~\cite{Koposov2010} which are shown as red points. The data agree well with the mock stream's angular position in the sky. The minor offset between the data and the simulated track of the stream is because the phase space coordinate of the progenitor from ref.~\cite{Bovy2016} was obtained from a fit in which the Milky Way potential was left to vary freely, whereas we have evolved the stream in a fixed \texttt{MWPotential2014}. This offset however does not affect how well we can distinguish between WDM and CDM.

\subsection{Modeling stream-subhalo impacts}
\label{subsec:subhaloimpacts}

A close encounter of a dark matter subhalo with a stellar stream imparts perturbations to the orbits of the stars in the stream which can be computed by the impulse approximation \cite{Yoon2011,Carlberg2013,Erkal2015,Sanders2016}. In this approximation, the subhalo-stream encounter is modeled as an instantenous velocity kick imparted to the stars in the stream at the point of closest approach. In ref.~\cite{Sanders2016} it was shown that subhalo-stream interactions can be efficiently modeled in frequency-angle space by transforming the velocity kicks, $\delta\vect{v}^{g}$, in frequency-angle space using the Jacobians $\partial\vect{\Omega}/\partial\vect{v}$ and $\partial\vect{\theta}/\partial\vect{v}$. The equations of motion for the stars in frequency-angle space before the subhalo impact are $\vect{\Omega} = \vect{\Omega}_{0} = $ constant and $\vect{\theta} = \vect{\Omega}_{0}t + \vect{\theta}_{0}$, where $(\vect{\Omega}_{0},\vect{\theta}_{0})$ is the frequency-angle coordinate when the star was stripped from its progenitor. If the  subhalo impacts at $t^{g}$, then the equations of motion of the star becomes $\vect{\Omega} = \vect{\Omega}_{0} + \delta\vect{\Omega}^{g}= $ constant and $\vect{\theta} = \vect{\Omega}_{0}t + \delta\vect{\Omega}^{g}(t - t^{g}) + \delta\vect{\theta}^{g} + \vect{\theta}_{0}$, where $\delta\vect{\Omega}^{g}$ and $\delta\vect{\theta}^{g}$ are the frequency and angle kicks corresponding to $\delta\vect{v}^{g}$. Following refs.~\cite{Sanders2016,Bovy2016a}, instead of computing $\delta\vect{v}^{g}$ over the full 6-dimensional phase space volume, we only compute along the 1-dimensional mean track of the stream at the time of impact. Moreover, the angle kicks, $\delta\vect{\theta}^{g}$ were shown to be small compared to the frequency kicks, $\delta\vect{\Omega}^{g}$ after one orbital period \cite{Sanders2016}, therefore in the context of the GD-1 stream, which is much older than its orbital period, we can neglect the angle kicks. Extensive tests of these approximations were performed in ref.~\cite{Bovy2016} and they concluded that the approximations work well at least in the subhalo mass range of interest here.

In order to simulate the effects of multiple subhalo impacts of different masses, at different times in the orbit and locations along the stream, we need a prescription for sampling multiple subhalo impacts. For this we follow the sampling procedure described in detail in section 2.3 of ref.~\cite{Bovy2016a}. The only difference in the WDM case is the Poisson sampling of the number of impacts of different masses. 

We consider the subhalos to follow a Hernquist profile, and use the relation for the scale radius, $r_{\rm{s}}$, from ref.~\cite{Erkal2016},
\begin{equation}
r_{\rm{s}} = 1.05 ~\rm{kpc} \left(\frac{M_{\rm{sub}}}{10^8 \Msun}\right)^{0.5},
\label{eq:rs}
\end{equation}
which was obtained for CDM subhalos with Hernquist profile by fitting the circular velocity - mass relation from the publicly available Via Lactea II catalogs \cite{Diemand2008}. We elaborate in appendix \ref{sec:scaleradius} on why we are justified in using the same $r_{\rm{s}}$ fitting formula for WDM subhalos. Less massive and smaller dark matter halos need to pass closer to a stellar stream compared to more massive and larger subhalos to result in an observable effect on the stream. To capture this effect in our simulations, we set the maximum impact parameter equal to five times the scale radius of the dark matter subhalo following ref.~\cite{Bovy2016a}. The maximum impact parameter is therefore smaller for lower mass subhalos and larger for more massive subhalos. To sample the velocity distribution of the DM subhalos, we use a Gaussian with a velocity dispersion of 120 km s$^{-1}$ which is the measured radial velocity dispersion within a galactocentric distance of 30 kpc for a collection of halo objects as found in ref.~\cite{Battaglia2005}. 

For a particular mass of thermal WDM, one can find the number of subhalos in a specific mass range within a spherical radius by integrating eq.~\eqref{eq:dndMw} over the subhalo mass range and the chosen radial range.
For the GD-1 stream generated in this work, the mean spherical radius of the stream is $\sim 23$ kpc. If dark matter constitutes of a thermal WDM of particle mass 3.3 keV (or a 7 keV sterile neutrino with lepton asymmetry parameter equal to 8.66), we find 0.47 subhalos in the mass range $[10^{6} - 10^{9}]~ \Msun$ within 23 kpc from the galactic center if subhalo disruption due to baryonic effects are not taken into account. By taking the 47\% reduction of number of subhalos due to baryonic effects, this number is reduced to $\sim 0.25$. As a comparison, in case of a 1 GeV CDM candidate in the same mass range and within the same radius, we find $\sim 57.78$ subhalos when no baryonic effect is included, and $\sim 30.6$ when we include baryonic effects. For the rest of this work we include the 47\% reduction of number of subhalos due to the effect of baryons in all our results.

\begin{figure}
\centering
\includegraphics[scale=0.5]{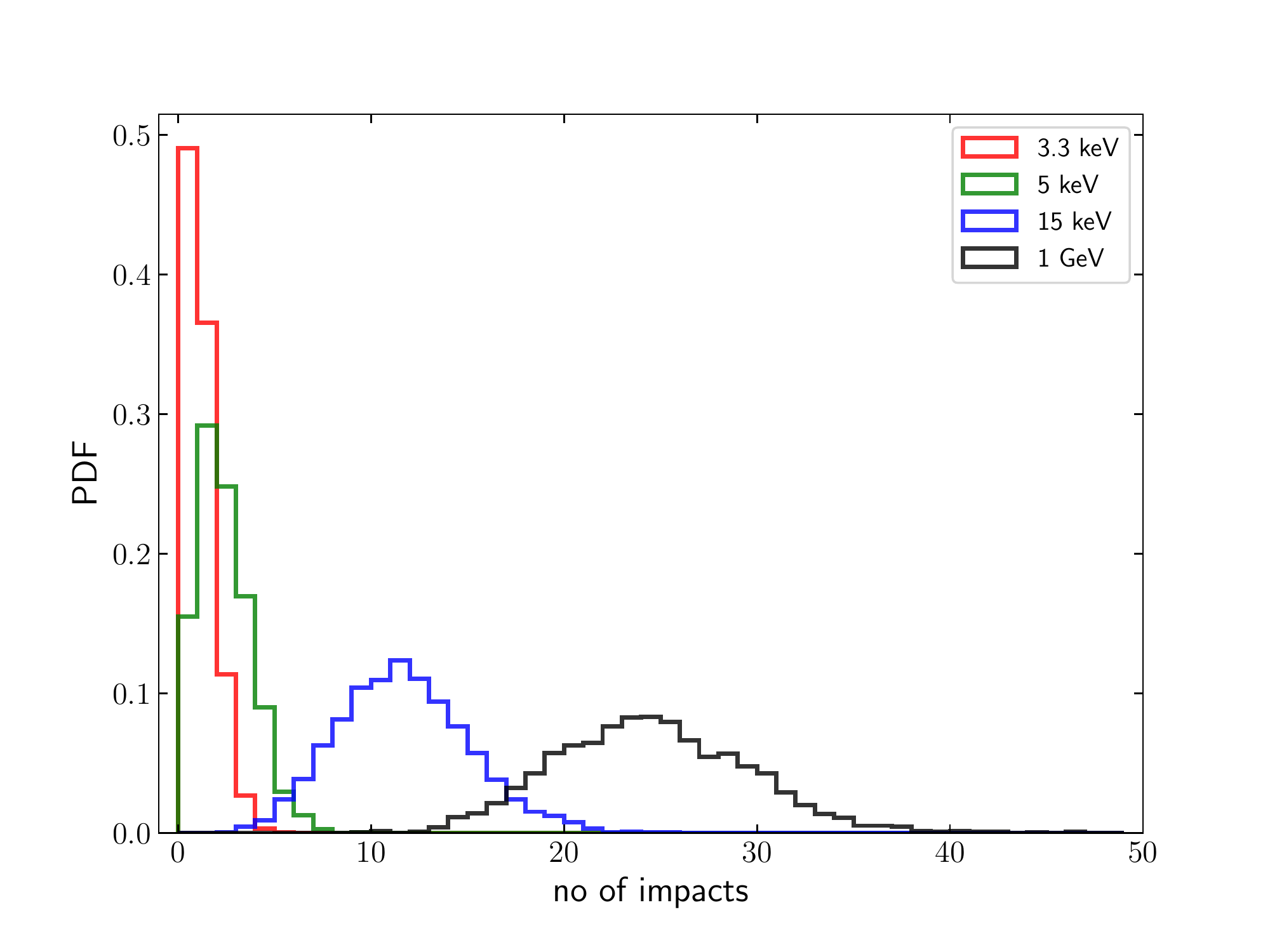}
\caption{PDF of the number of impacts that a GD-1 like stream had over 9 Gyr in different cases of dark matter particle mass. Each PDF was constructed out of 2100 simulations.}
\label{fig:impact_pdf}
\end{figure}

 The expected number of impacts over the age of a stream depends on the number density of subhalos as well as on the impact parameter (see eq.~(1) in ref.~\cite{Bovy2016a}). In the WDM case, there are fewer subhalos compared to the CDM case, and as a result the probability of stream-subhalo encounter is much smaller if dark matter is warm. This is shown in figure \ref{fig:impact_pdf} for different cases of dark matter particle mass. Each impact PDF is constructed out of 2100 simulations of the GD-1 stream evolved in different dark matter particle mass scenarios. 

 To find the impact rates, we compute the number of subhalos within 23 kpc in each mass decade by integrating eqs.~\eqref{eq:dndMc} and \eqref{eq:dndMw}. We determine the number of impacts $N$ over the entire mass range of subhalos in the WDM scenario by sampling the number of impacts from a Poisson distribution for the expected number of impacts. We then sample $N$ masses from the impact-mass distribution. Note that this distribution differs from eq.~\eqref{eq:dndMw}, because it needs to account for the shrinking maximum impact parameter as the subhalo mass decreases and vice versa. All other parameters are then sampled exactly as in the CDM case considered in ref.~\cite{Bovy2016a}. The total rate of impact is 0.72 for a 3.3 keV thermal WDM, 1.95 for a 5 keV thermal WDM, and 24.8 for a 1 GeV CDM particle. These rates are calculated by assuming that the rate of impact for each mass is that corresponding to its mass decade and is computed at the center of the logarithmic mass bin.  In the subhalo mass bins $[10^{6} -  10^{7}]~ \Msun$, $[10^{7} -  10^{8}]~ \Msun$, and $[10^{8} -  10^{9}]~ \Msun$, the impact rates for the 3.3 keV WDM scenario are $\sim 0.04$, $\sim 0.17$ and $\sim 0.49$, for the 5 keV WDM are $\sim 0.17$, $\sim 0.6$, $\sim 1.18$ and for the 1 GeV CDM scenario they are $\sim 15.94$, $\sim 6.34$, and $\sim 2.53$, respectively.

In order to efficiently calculate the stream-subhalo impacts, we use the \emph{line-of-parallel-angle} approach as explained in ref.~\cite{Bovy2016a}. In this approach, multiple impacts occurring at the same time do not increase the computational cost. Therefore, the impacts are allowed to happen at a set of equally spaced discrete times along the past orbit of the stream. In order for the structure of the stream not to be affected by the discrete time sampling, the sampling needs to be high enough. Furthermore, sampling impacts more frequently than the radial period of the stream allows us to explore stream-subhalo encounters at different epochs of the streams orbit. In Appendix C of ref.~\cite{Bovy2016a} it was shown that the statistical properties of a perturbed stream converge for greater than 16 different impact times. We considered 64 different impact times for the 9 Gyr stream. This would amount to a time interval of $\sim 140$ Myr which is shorter than the radial period of $\sim 400$ Myr for the GD-1 stream. 
\section{Results}
\label{sec:results}

\subsection{Mock perturbed GD-1 stream}

Using the above mentioned machinery for generating perturbed stellar streams, we carry out 2100 simulations each for the 3.3 keV and 5 keV thermal WDM and 1 GeV CDM scenarios for a GD-1 like stream, considering subhalo impacts in the mass range of $[10^6 - 10^9]~\Msun$. For each simulation, we transform the density of the stream from frequency-angle space to configuration space $(\phi_{1},\phi_{2})$ and apply angular cuts to match the observed extent of the GD-1 stream. As evident from figure \ref{phi12}, the angular extent of the stream along $\phi_{2}$ is very small, therefore for the rest of this work, we will only analyze the stream as a function of $\phi_{1}$. 

\begin{figure}[t]
    \centering

    \begin{minipage}{0.5\textwidth}
        \centering
        \includegraphics[width=\linewidth]{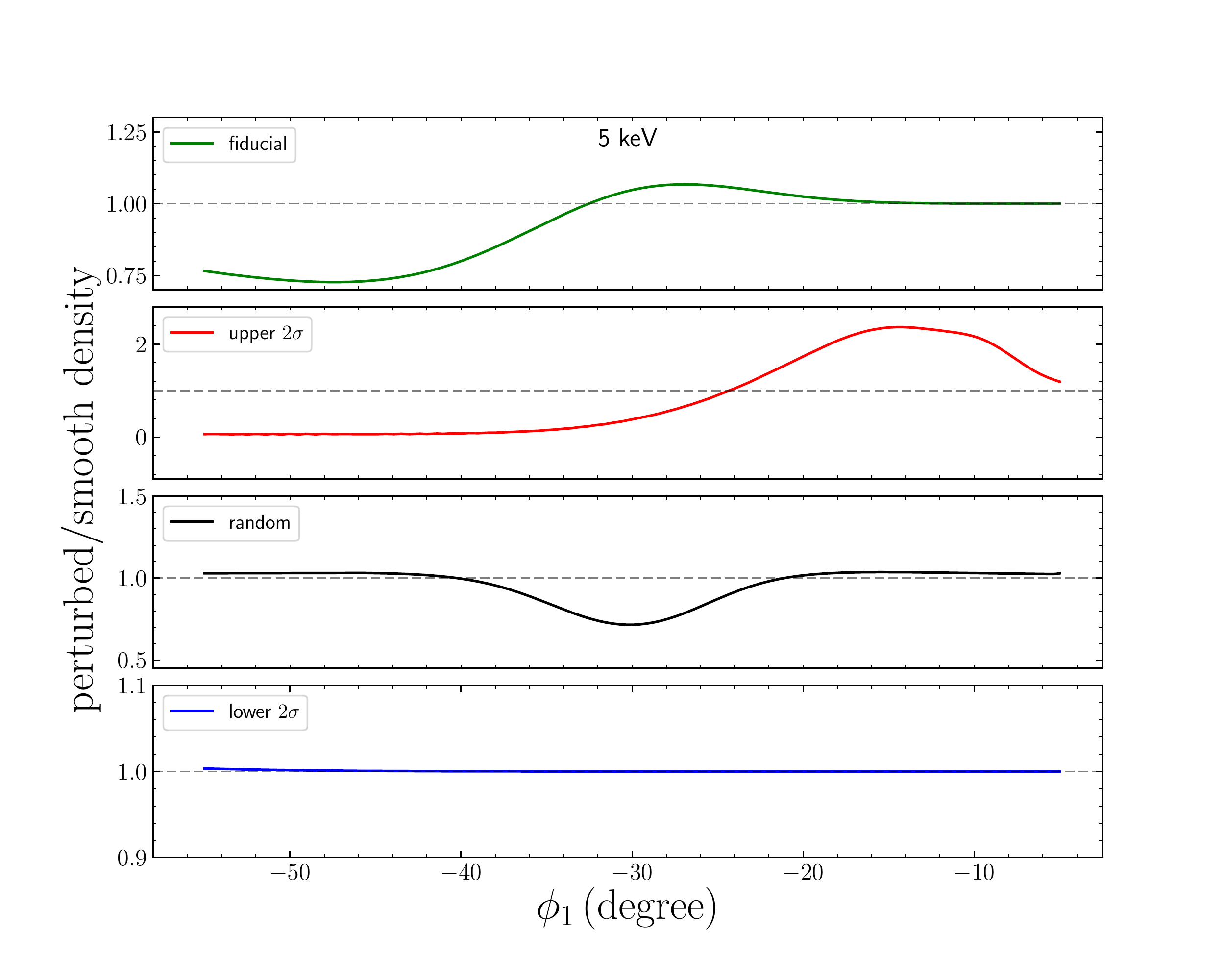}

    \end{minipage}%
    \begin{minipage}{0.5\textwidth}
        
        \includegraphics[width=\linewidth]{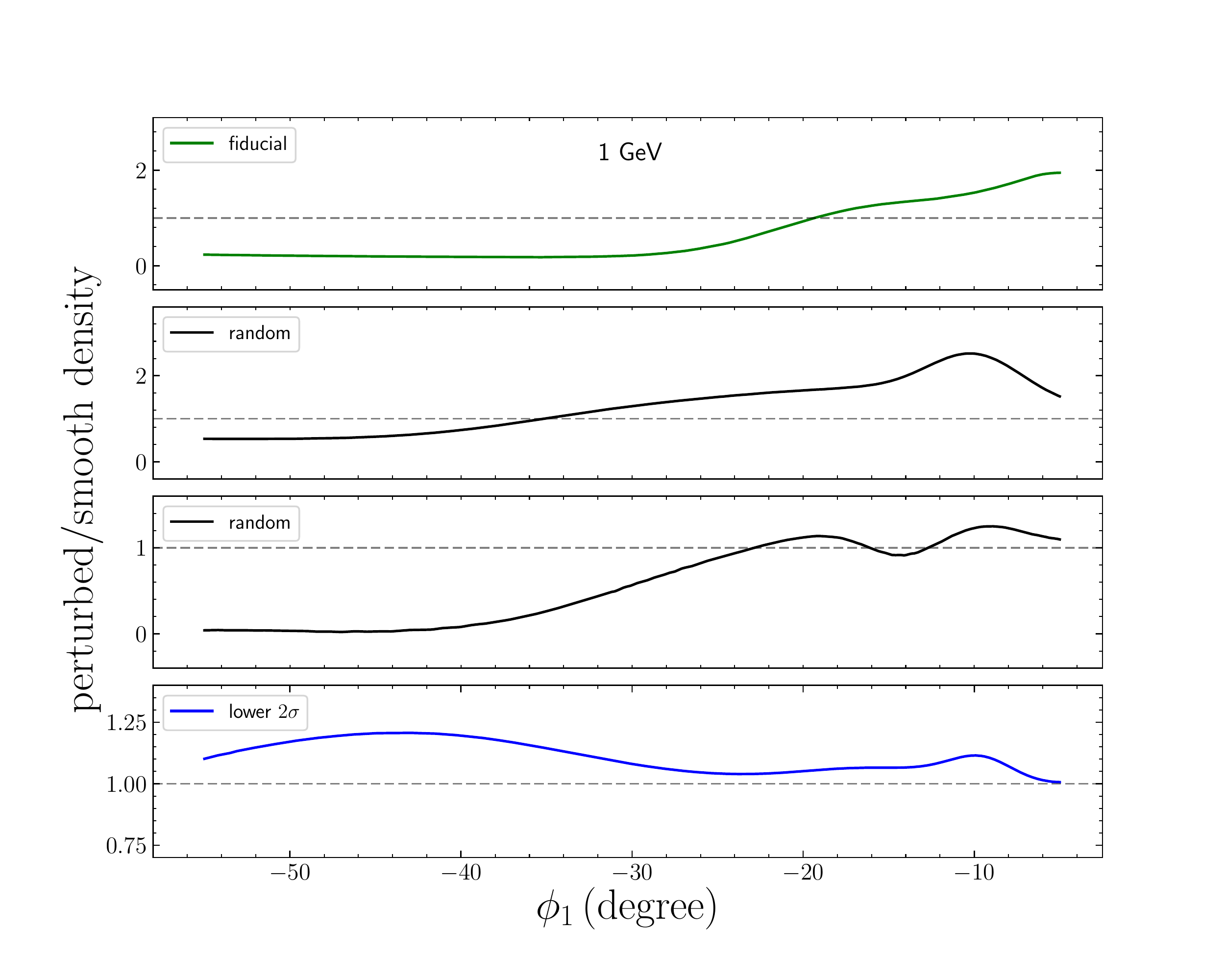}

    \end{minipage}
\caption{Density contrast of four different realizations of the perturbed GD-1 stream. The left panel shows the case in which dark matter is composed of thermal WDM with particle mass 5 keV and the right panel shows the case for a 1 GeV CDM candidate. The black curves show randomly chosen density contrasts. The green curves denote density contrast for the fiducial cases which we treat as mock observed data and are analyzed for inferring the mass of dark matter in section \ref{sec:infer}. The red curve indicates a realization whose power is close to the upper $2\sigma$ bound of the power spectrum dispersion, while the blue curves indicate the realizations that are close to the lower $2\sigma$ bound. In the 5 keV case, this happens when the stream had no subhalo impact. See section \ref{sec:outlier} for discussion on these extreme cases. }
\label{fig:denscont}
\end{figure}

\begin{figure}[t]
\centering
    \includegraphics[scale=0.49]{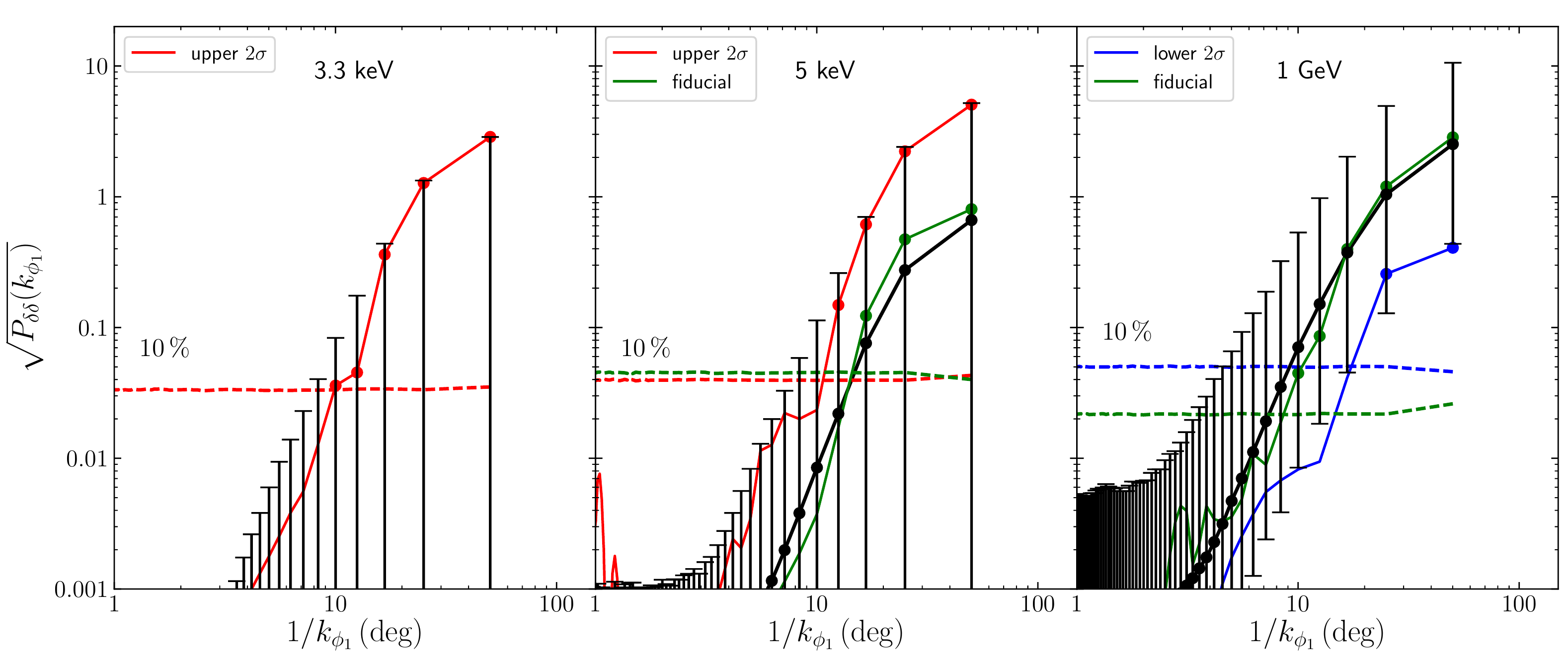}

    \caption{Power spectrum of the density contrast of the mock GD-1 stream in the 3.3 keV WDM (left panel), 5 keV WDM (middle panel) and 1 GeV CDM (right panel) scenarios. The black solid line is the median power of the 2100 simulations. For the 3.3 keV case, the median power is close to 0 and hence not shown. The error bars represent the $2\sigma$ (95\%) spread of the power spectra of the 2100 stream density simulations. The green solid curves (wherever shown) represent the power spectrum of the fiducial cases whose density contrasts are shown in figure \ref{fig:denscont}. The red and blue curves (wherever shown) represent the power for cases close to the upper and lower $2\sigma$ bound of the dispersion in power spectrum, respectively. The dashed curves indicate the median power corresponding to 10\% noise in the density and their color represents the case they correspond to. The colored dots specify the angular scales at which the power is above the noise floor.} 
    \label{fig:Pk_disp}
\end{figure}

In figure \ref{fig:denscont} we show the density contrast (density of the perturbed stream divided by that of the unperturbed stream) of four different cases from the 2100 simulations for the 5 keV and 1 GeV scenarios. Black curves denote randomly chosen density contrasts, while green curves denote density contrasts for cases whose power is close to the median power of the 2100 simulations as shown in figure \ref{fig:Pk_disp}. We treat the cases shown in green as fiducial cases which are used to constrain the mass of dark matter in section \ref{sec:infer}. The red curve in the 5 keV WDM scenario in the left panel of figure \ref{fig:denscont} corresponds to a case whose power is close to the upper $2\sigma$ bound of the power spectrum dispersion, while the blue curves in both panels of the same figure correspond to cases whose power is close to the lower $2\sigma$ bound as shown in figure \ref{fig:Pk_disp}. For the 5 keV WDM, the case close to the lower $2\sigma$ bound occurs for a stream that had no subhalo impacts, and as a result the density contrast is equal to one. We do not consider a case in the 1 GeV CDM scenario whose power is close to the upper $2\sigma$ bound of the power spectrum dispersion because as we will discuss in section \ref{sec:infer}, our method can not distinguish between dark matter models if the mass of dark matter is greater than $\sim 15$~keV. In section \ref{sec:outlier}, we investigate these cases of extreme power and how they can bias our analysis.

In our stream simulation, the progenitor emits new stars constantly. In scenarios where the stream had many impacts, the bulk of the stream gets disrupted and stars are pushed towards the extremities. In such cases, the stars which are pushed towards the progenitor add up to the new stars which have recently been emitted from the progenitor, causing a large overdensity near the location of the progenitor ($\phi_{1}=0^{\circ}$). Such large overdensities are somewhat unphysical since in a real stream some stars pass the progenitor, moving from one arm of the stream to the other, and our simulations ignore this effect. In order to remove this potentially artificial peak, we cut the stream at $\phi_{1} > -5^{\circ}$. This choice of the $5^{\circ}$ cut removes the large peaks close to the progenitor, while leaving the smaller physical peaks which are expected due to the disruption of the subhalo.

For each case we binned the density in $ 0.1^{\circ}$ $\phi_{1}$ bins. If we assume that LSST can go below the main-sequence turn-off point and accurately resolve almost all the member stars of the GD-1 stream which has $\approx1,000\,\mathrm{stars\, deg}^{-1}$ \cite{Koposov2010}, then this would correspond to a shot noise of 10\% for a bin width equal to $0.1^{\circ}$. Therefore, we have considered a constant noise of 10\% of the density contrast in each $\phi_{1}$ bin.

We compute the 1-dimensional power spectrum of the density contrast of the stream $P_{\delta\delta}(k_{\phi_1}) = \langle \delta(k_{\phi_1})^2 \rangle$ using the \texttt{csd} routine in \texttt{scipy} \cite{Jones2001c} and do not divide by the sampling frequency. In the power spectrum plots, we have plotted the square root of the power along $y$-axis and inverse of the frequency (1/$k_{\phi_1}$) along $x$-axis. In this form, the power spectrum plots represent the power at a particular angular separation which is indicative of the correlation of the density contrast at that angular scale. In figure \ref{fig:Pk_disp}, we show the power spectrum of the density contrast of the GD-1 stream for the 3.3 keV and 5 keV WDM, and the 1 GeV CDM scenarios. The black solid curves show the median power spectrum of the 2100 simulations. For the 3.3 keV case, owing to very few impacts in majority of the simulations the median power is close to 0. The error bars show their $2\sigma$ dispersion around the median. The green solid curves show the power spectrum of our fiducial cases whose density contrasts are shown in figure \ref{fig:denscont}.  The red and blue solid curves show a case close to the upper and lower $2\sigma$ bound of the power spectrum dispersion, respectively. In both the 3.3 keV and 5 keV WDM scenarios the lower bounds are close to zero. This is again due to either no subhalo impacts or impacts that did not result in significant density fluctuations. The dashed lines show the 10\% noise power and their color represents the case they correspond to. The noise power spectrum is the median of 10,000 Gaussian noise realizations of the stream density and therefore independent of the density bin width. For the fiducial 1 GeV case, the noise power is much lower than the other cases since the bulk of the stream is pushed away due to subhalo encounters leaving the density along most of the stream close to 0 (top right panel figure \ref{fig:denscont}). The density noise being 10\% of the density has therefore very low power. The colored dots represent the angular scales at which the power is above the corresponding noise floor and hence are used in the ABC analysis to infer the dark matter particle mass (see section \ref{sec:infer}). 

As evident from figure \ref{fig:Nsub}, there are many more subhalos in the mass range $[10^6 - 10^9]~\Msun$ in the 1 GeV case compared to the 5 keV and 3.3 keV cases, which leads to a higher chance of stream-subhalos encounters as suggested in figure \ref{fig:impact_pdf}, resulting in higher density fluctuations along the stream in the 1 GeV case compared to the 5 keV and 3.3 keV cases. This is easily seen from the density contrasts plotted in figure \ref{fig:denscont}. 

 The difference in the density fluctuations translates to the power spectrum and hence explains the more power at the largest scales in the 1 GeV case compared to the 5 keV case. The wide dispersion in the power spectra in either case is due to the range of possible ways the stream-subhalo impacts can occur resulting in different levels of density fluctuations along the stream. 
 Comparing the 5 keV and 1 GeV cases in figure \ref{fig:Pk_disp}, it is clear that there is more power at smaller scales in the 1 GeV case. This is because if dark matter is cold then there are more low mass subhalos whose impacts with the stream give rise to density fluctuations at smaller scales. On the contrary, if dark matter is warm then as discussed in section \ref{sec:WDM}, substructures less massive than the half-mode mass are strongly suppressed. As a result, majority of the impacts are by the more massive subhalos, giving rise to large scale density fluctuations. The noise power appears flat since we considered a constant noise level throughout the stream. Power at scales below the noise floor do not convey any useful information.

\subsection{Inferring the dark matter particle mass}
\label{sec:infer}

\begin{figure}[t]
\centering
\includegraphics[scale=0.5]{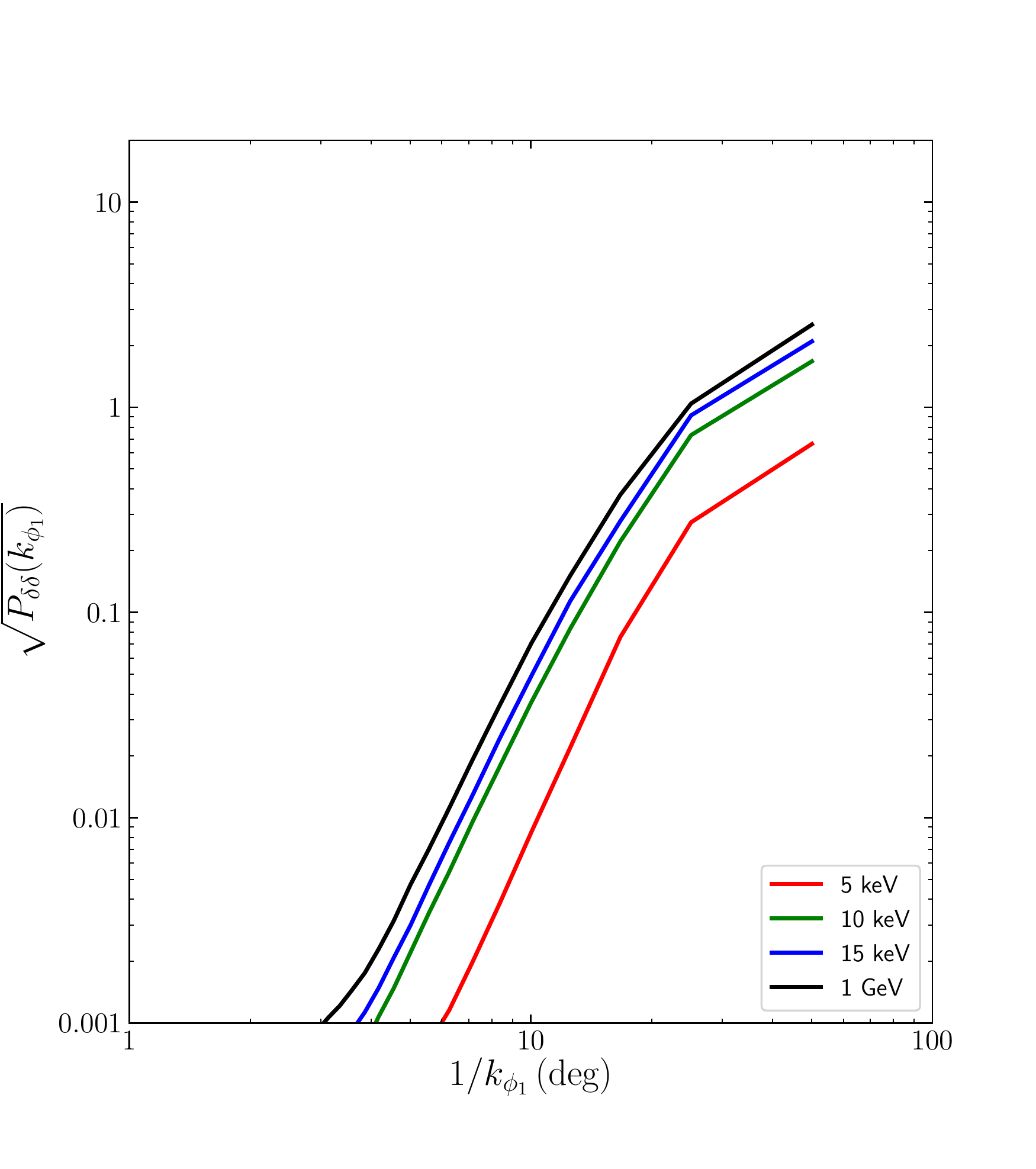}
\caption{Power spectrum of the median of 2100 simulations of the GD-1 stream in different WDM particle mass scenarios. The median power spectra for WDM particle mass larger than $\sim 10$~keV tend to overlap, and hence it is no longer possible to distinguish at high significance between WDM models that have particle mass greater than 10 keV. The power spectrum of cases with WDM particle mass greater than 15 keV is very close to the 1 GeV CDM case and therefore WDM models with particle mass greater than $\sim 15$~keV can not be distinguished from the CDM model.} 
\label{fig:Pk_conv}
\end{figure}

In this section we use the statistical properties of gaps in a stellar stream to distinguish between CDM and WDM. To achieve this, we use the power spectrum of the density contrast of the perturbed stream to constrain the mass of the dark matter particle. We employ the ABC method to construct an approximate posterior probability distribution function (PDF) of the mass of the dark matter particle. The ABC method is a likelihood-free approach of Bayesian parameter inference in which an approximate posterior PDF of the parameters in the problem is constructed using simulator outputs and by comparing the outcome with observed data. We run the simulations by randomly drawing the mass of the dark matter particle from a prior distribution of the mass, which we have assumed to be a uniform distribution between [0.1,16] keV. The upper limit of the prior was picked based on the result presented in figure \ref{fig:Pk_conv}, which shows that the median power spectrum for cases with mass of WDM greater than 15 keV converge and are indistinguishable from each other. This is consistent with figure \ref{fig:Nsub}, which shows that increasing the mass of the WDM particle shifts the subhalo mass function close to the CDM case in the subhalo mass range $[10^6 - 10^9]~\Msun$. 

 The ABC works by accepting those simulations which are within some pre-defined tolerance of the data summaries. As our data summaries we have taken the power at the three largest observed scales in the 5 keV case and the five largest observed scales in the 1 GeV case, since those scales are above the noise floor of the fiducial cases (see figure \ref{fig:Pk_disp}). We do not consider the 3.3 keV case here, since as seen in figure \ref{fig:Pk_disp} the median power is close to zero for this case. Additionally, we do not consider power at scales below the noise floor because they are very noisy and thus are incapable of discriminating between different WDM models. The level of noise in the data is therefore crucial for our method. The tolerances are made as small as it is allowed by the noise in the data. In order to sample the entire range of the prior distribution properly, we ran $\sim 120,000$ simulations. Since the most time consuming part of the ABC approach is running these simulations, we follow the same strategy as adopted in ref.~\cite{Bovy2016a}, i.e.~for each WDM mass we produce 100 simulations by adding 100 realizations of the noise. Therefore, effectively, we ran $\sim 12,000,000$ simulations. 

\begin{figure}[t]
    \centering
    \begin{minipage}{.5\textwidth}
        \centering
        \includegraphics[width=\linewidth]{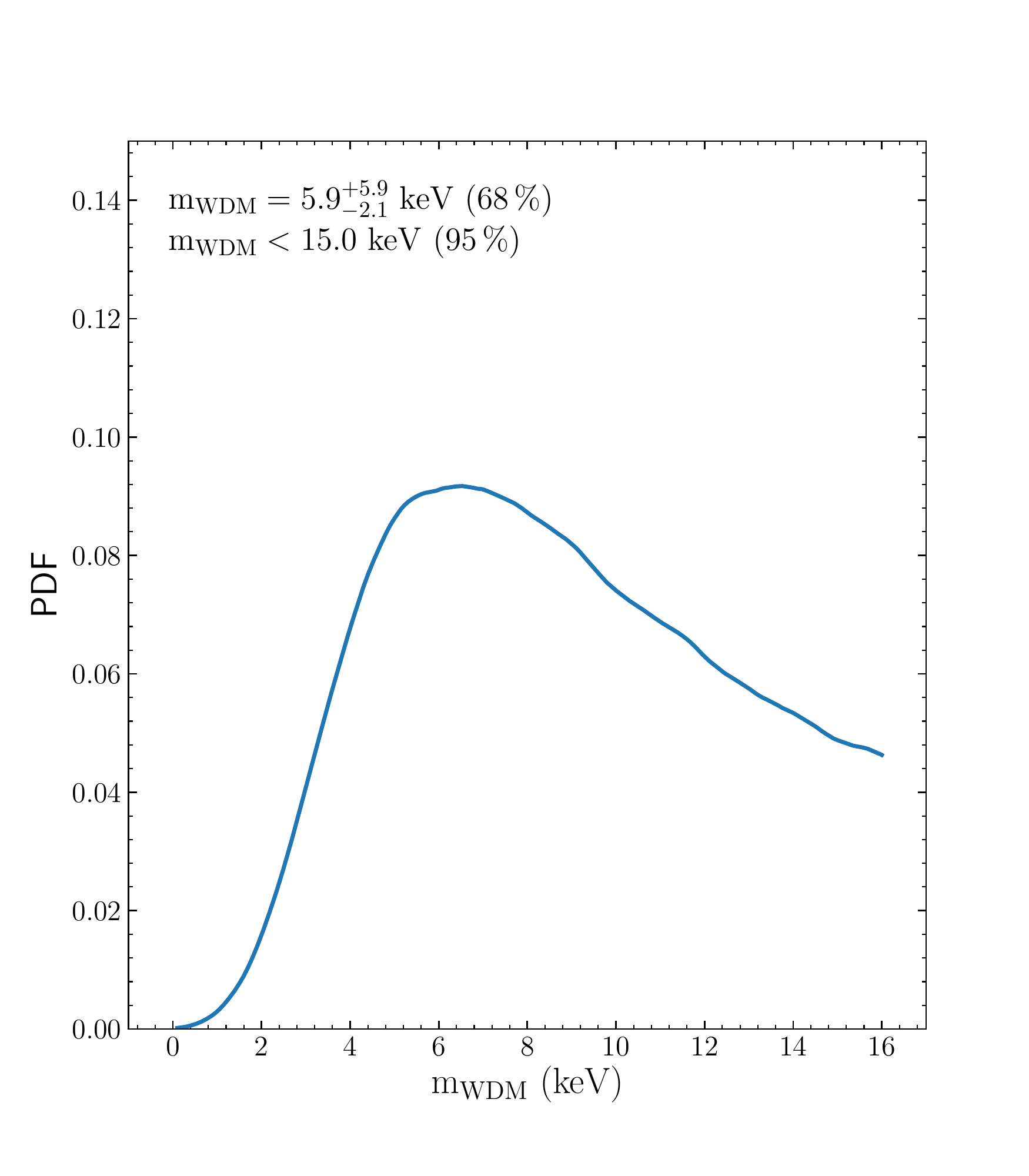}

    \end{minipage}%
    \begin{minipage}{0.5\textwidth}
        \centering
        \includegraphics[width=\linewidth]{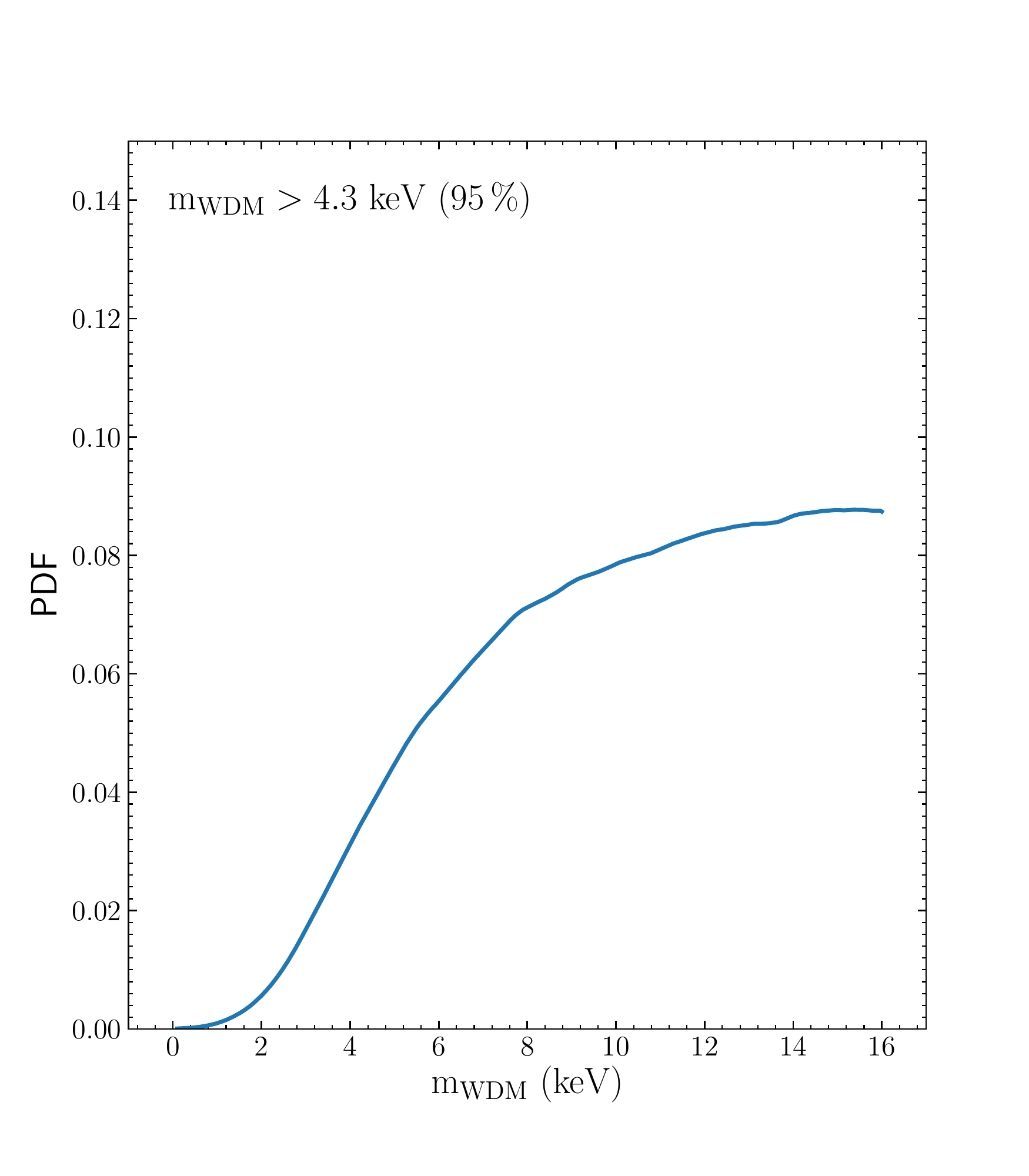}
        
    \end{minipage}
    \caption{Posterior PDFs of the mass of dark matter obtained by running simulations of the GD-1 stream and using the ABC method to match the power at the largest scales above the noise floor (shown by red dots in figure \ref{fig:Pk_disp}) to the fiducial cases. The left panel shows the posterior PDF in the 5 keV WDM scenario and the right panel shows the posterior PDF in the 1 GeV CDM scenario. For these typical cases the ABC is able to distinguish between WDM and CDM and also constrain the mass of WDM. }
        \label{fig:PDF}
\end{figure}

Figure \ref{fig:PDF} shows the posterior PDF of the dark matter particle mass obtained by running simulations of the GD-1 stream and using the ABC method as explained above in the 5 keV WDM and the 1 GeV CDM scenarios. For the 5 keV case, the PDF peaks at $5.9^{+5.9}_{-2.1}$ keV (68\% confidence) and the 95\% upper limit is 15.0 keV. In the 1 GeV CDM case, the PDF plateaus indicating the true dark matter mass is beyond the upper limit of the prior. The 95\% lower limit is 4.3 keV.

\begin{figure}[!htb]
    \centering
    \begin{minipage}{.5\textwidth}
        \centering
        \includegraphics[width=\linewidth]{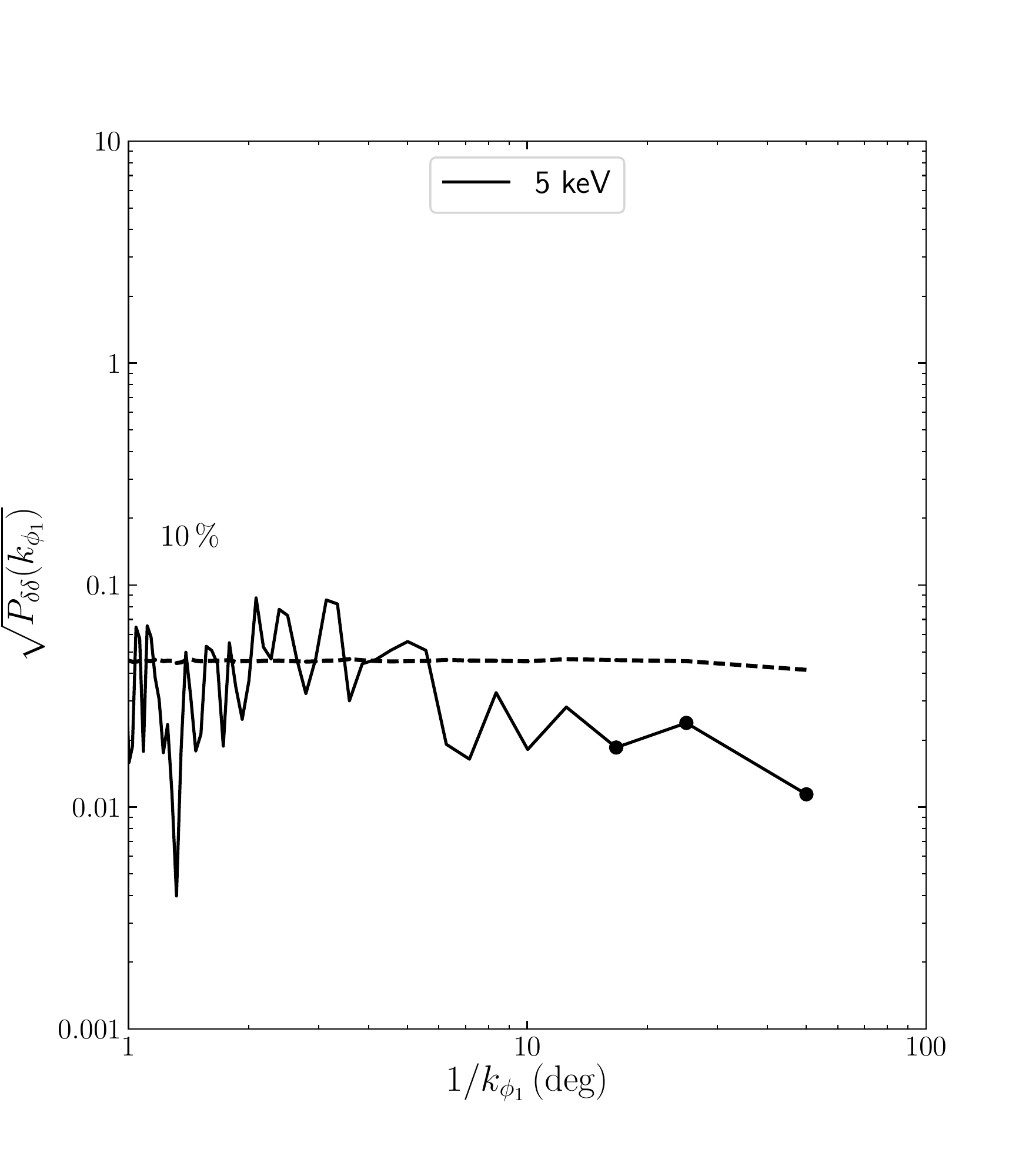}
        
    \end{minipage}%
    \begin{minipage}{0.5\textwidth}
        \centering
        \includegraphics[width=\linewidth]{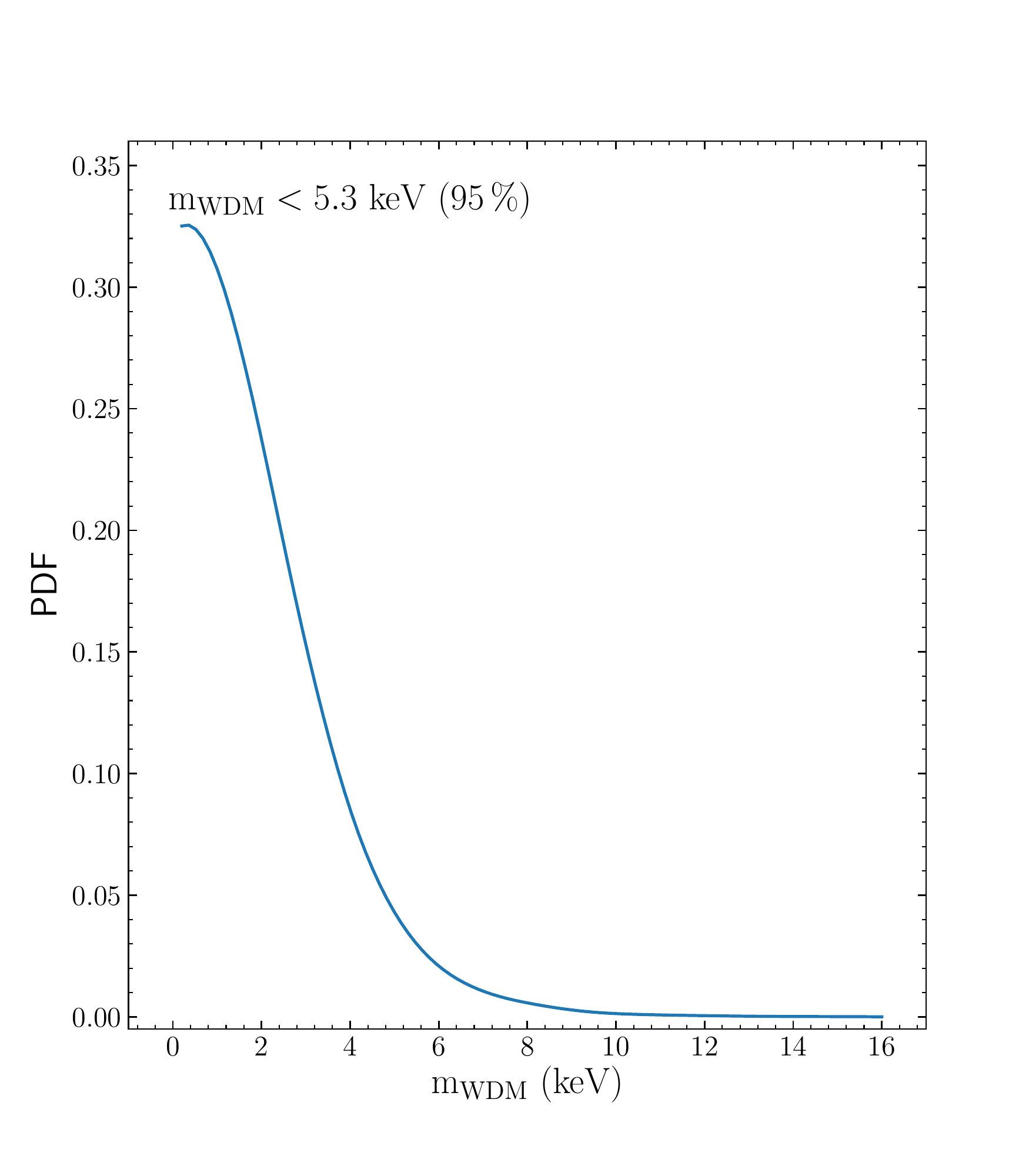}

    \end{minipage}
    \caption{Figure demonstrating the case in which the intrinsic power of the stream density is below the noise floor, as a result of either no subhalo impact or impacts that did not contribute to any significant density fluctuations. Left panel: One realization of the noise power (solid black line) which we treat as our mock measurement of the stream density power spectrum, and the 10\% noise floor (dashed black line). The black dots represent the data summaries which are the power at the three largest scales. Right panel: Posterior PDF of the dark matter mass obtained using the ABC method.}
        \label{fig:noise_PDF}
\end{figure}

Next, we investigate the case in which the intrinsic power of the stream is below the noise floor. As shown in the power spectrum dispersion plots in figure \ref{fig:Pk_disp}, such cases may arise when dark matter is warm and the stream suffers few or no subhalo impacts (e.g., the density contrast for the 5 keV case corresponding to the lower $2\sigma$ bound).
In such cases the measured power is dominated by noise. The left panel of figure \ref{fig:noise_PDF} shows one randomly picked realization of the noise in the 5 keV case which we use as the measured power. We consider the power at the three largest scales to be data summaries. Since the power at all these three scales are below the median noise floor, the ABC accepts any simulation whose power at those scales are below the noise floor. The resulting posterior PDF of the dark matter mass is shown in the right panel of figure \ref{fig:noise_PDF}, which demonstrates that the 95\% upper limit on the mass of WDM is 5.3 keV. This is consistent with the fact that for a lower mass WDM, there are fewer subhalos and hence lower chances of stream-subhalo encounters, compared to the case of a higher mass WDM.  

\section{Risks of using only one stream}
\label{sec:outlier}

\begin{figure}[t]
\centering
\begin{minipage}{.333\textwidth}
\centering
\includegraphics[width=\linewidth]{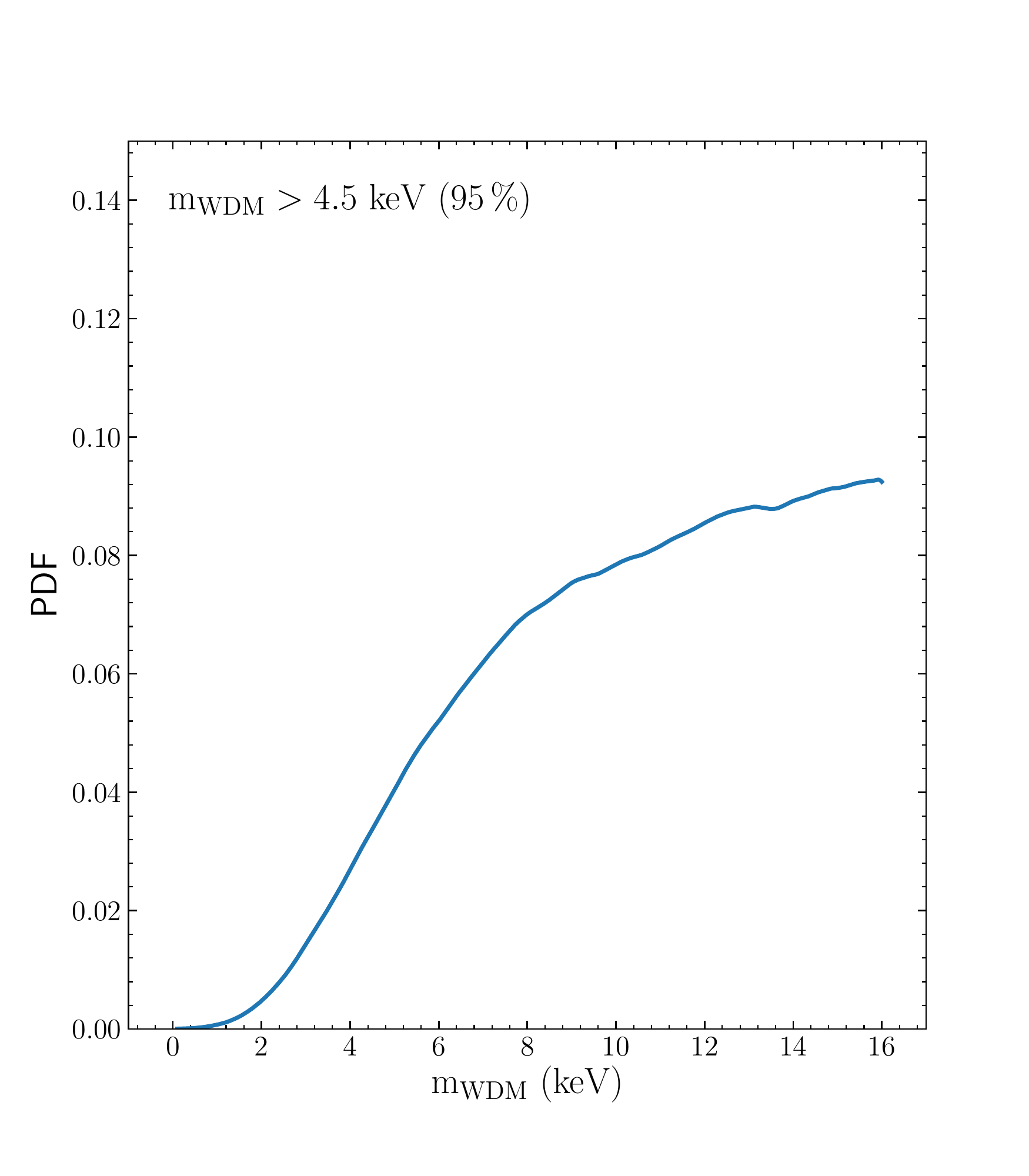}
\end{minipage}%
\begin{minipage}{0.333\textwidth}
\centering
\includegraphics[width=\linewidth]{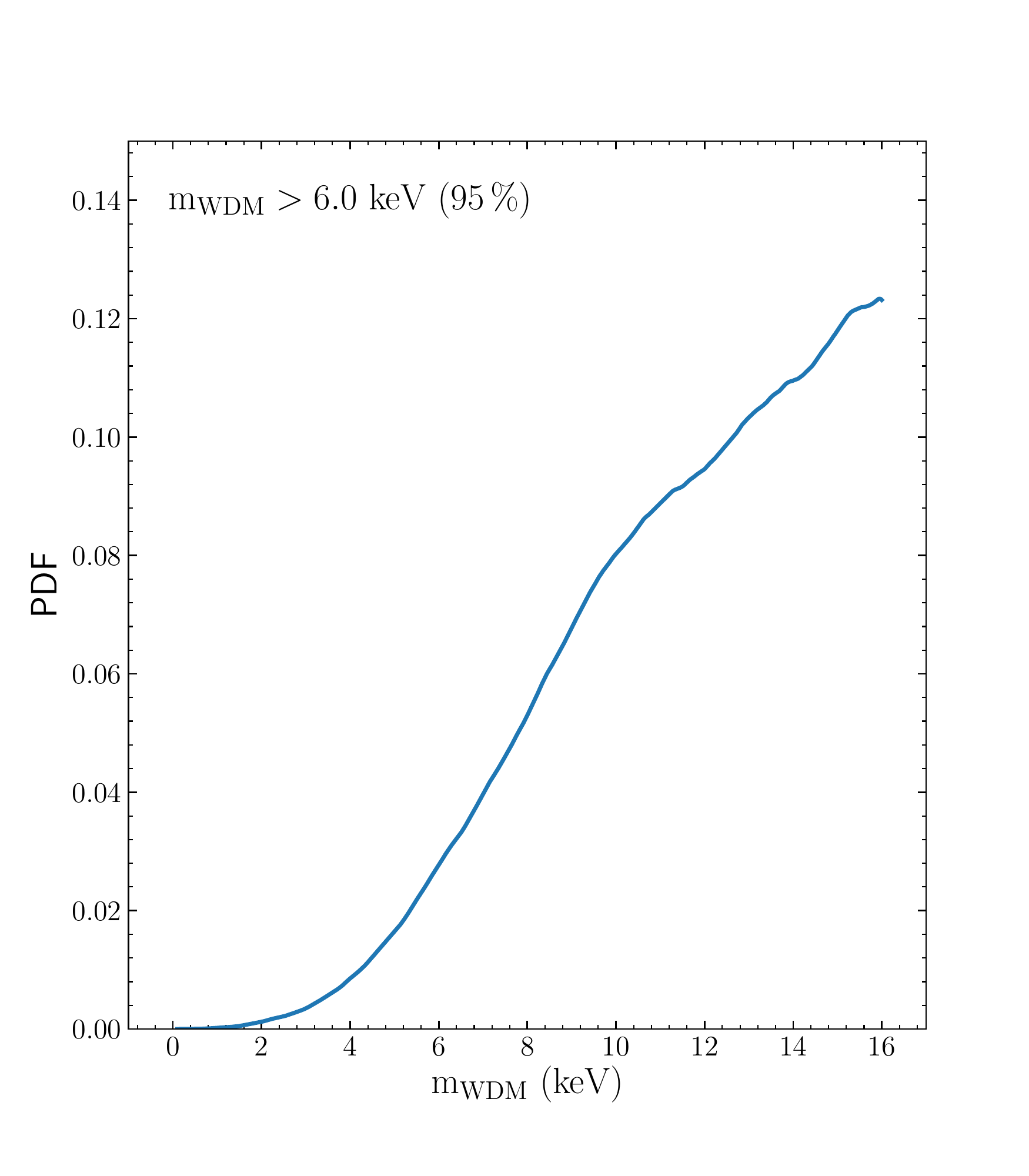}
\end{minipage}%
\begin{minipage}{0.333\textwidth}
\centering
\includegraphics[width=\linewidth]{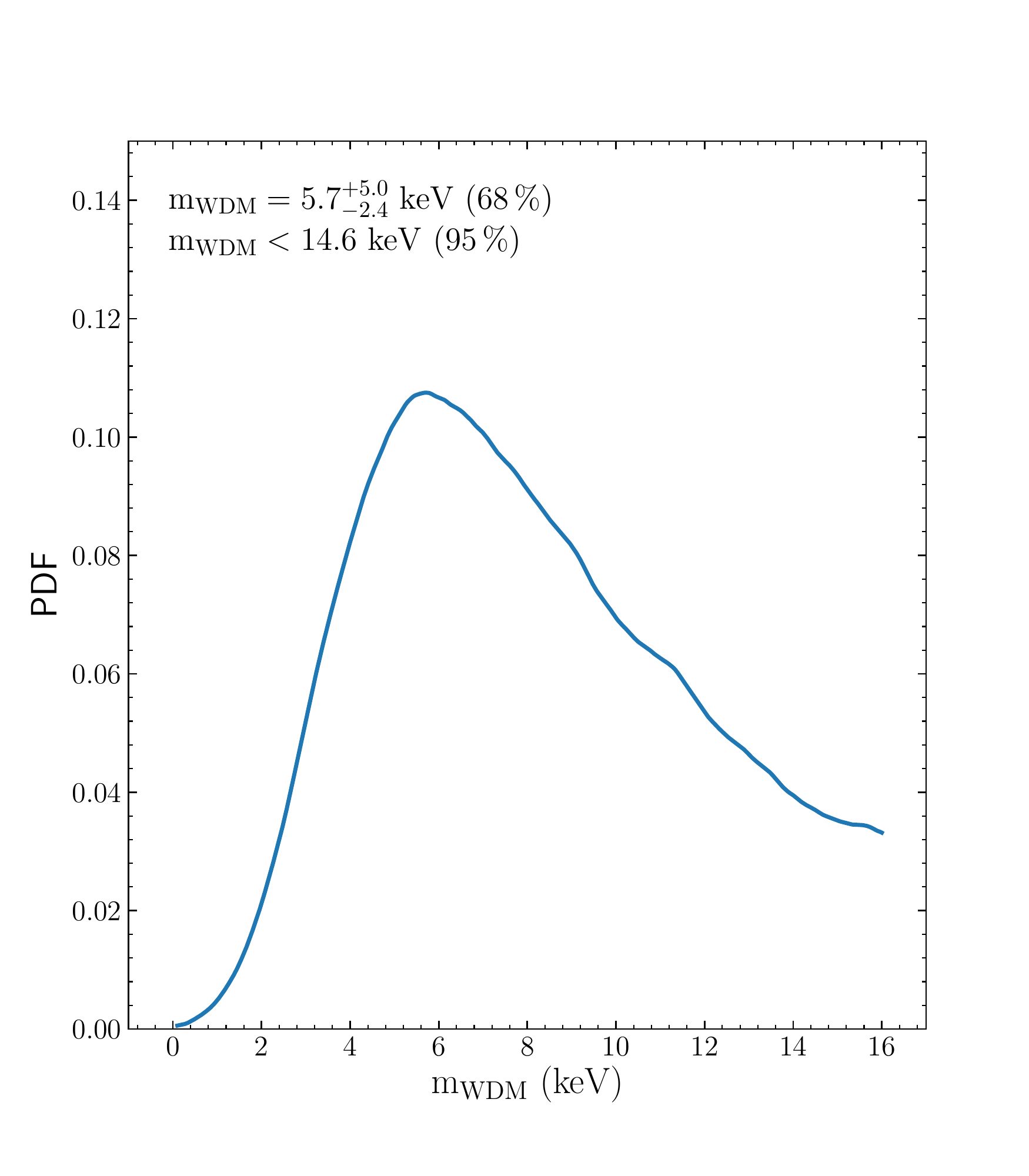}
\end{minipage}
    \caption{Posterior PDFs for the extreme cases. Left panel: upper $2\sigma$ case in the 3.3 keV scenario; middle panel: upper $2\sigma$ case in the 5 keV scenario; right panel: lower $2\sigma$ case in the 1 GeV scenario.}
        \label{fig:extreme_cases}
\end{figure}

In this section we discuss the perils of using only one stream for constraining the mass of WDM using the stream density power spectrum method presented above. As evident from figure \ref{fig:Pk_disp}, a stream-subhalo interaction can take place in a wide range of possible ways giving rise to the dispersion in the power spectrum. Therefore, it is not unexpected that we could end up with a stream that had either too few or too many subhalo hits resulting in its power spectrum to be $> 2\sigma$ away from the median. Such cases would result in a posterior that would rule out the true dark matter particle mass by $> 2\sigma$. We demonstrate this with three examples here. In the 1 GeV CDM scenario we pick a density simulation whose power spectrum is close to the lower bound of the $2\sigma$ dispersion, and for each 3.3 keV and 5 keV WDM scenarios we pick one realization with a power close to the upper $2\sigma$ bound. These cases are shown by the blue and red curves in figure \ref{fig:Pk_disp}. Treating them as mock data and considering a 10\% density noise, we carry out the ABC steps with power at the two largest scales for the 1 GeV case, five largest scales in the 3.3 keV case and four largest scales in the 5 keV case as data summaries to construct the posterior PDF for the mass of dark matter. The results are shown in figure \ref{fig:extreme_cases}. The posterior PDF predicts, in the 3.3 keV case (left panel), a lower 95\% limit on the mass of dark matter to be $4.5$ keV, in the 5 keV case (middle panel), a lower 95\% limit on the WDM particle mass to be $6$ keV and in the 1 GeV CDM case (right panel), the PDF constrains the dark matter particle mass to $5.7^{+5.0}_{-2.4}$ keV at 68\% and sets a 95\% upper limit at $14.6$ keV.

These results suggest that we can not rule out CDM or WDM simply based on the $2\sigma$ results of one stream. This shortcoming can be improved by using multiple stellar streams to constrain the mass of dark matter. This will not only remove the effects of the outlier cases but will also make the constraints more robust. We leave this analysis for a future publication.

\section{Conclusions}
\label{sec:conclusions}

In this paper we presented a methodology of investigating the particle nature of dark matter by analyzing the statistical properties of density fluctuations, in the form of power spectrum, induced in a stellar stream as a result of its gravitational encounter with dark matter subhalos. We accomplished this by exploiting the fact that if dark matter is warm then there will be fewer subhalos in the Milky Way host halo compared to the CDM case. This difference will result in a different rate of stream-subhalo encounters and will be apparent in the intrinsic power spectrum of the stream density. By using fast simulations of stream density in frequency-angle space we generated mock GD-1 stream data and exposed it to subhalo encounters in a 3.3 keV WDM, 5 keV WDM and 1 GeV CDM scenario. Due to the higher rate of impacts in the CDM scenario, we found that the stream density power spectrum has more intrinsic power in the CDM scenario compared to the WDM case. 

Assuming that LSST will be able to resolve most of the member stars of the GD-1 stream without contamination, we applied LSST-like noise to our mock data. With \textit{Gaia}'s magnitude threshold of 20, it will not provide much direct data on GD-1  for the purpose of our study. Nevertheless, Gaia will improve our knowledge of the Milky Way's gravitational potential in general and of the orbits of streams, which will be pivotal for our method. 

We used an ABC technique to perform rigorous inference on the dark matter mass using mock streams for the 5 keV WDM and the 1 GeV CDM fiducial cases. 
In the WDM case when the intrinsic power of the stream density is greater than the noise, we constrained the dark matter mass to $5.9^{+5.9}_{-2.1}$ keV (68\% confidence) and $< 15$ keV at 95\%. In the CDM scenario we found the lower limit on the mass of dark matter to be $4.3$ keV at 95\% confidence. These results indicate that with our methodology we can not only distinguish between CDM and WDM but also constrain the mass of WDM, if it's a few keV.

Because of the paucity of subhalos in the WDM scenario, it is possible that a stream doesn't accumulate enough density fluctuations in its lifetime so that its intrinsic power is less than the noise. In such cases, we obtained an upper limit on the mass of WDM to be 5.3 keV at 95\% confidence. We also explored three outlier cases of stream density perturbations in section \ref{sec:outlier} which can seriously limit our method's ability to distinguish between WDM and CDM. We stress that this situation can be ameliorated by applying our method on multiple streams.

\subsection*{Acknowledgments}

We thank Mark Lovell for many useful discussions and for providing us with the data from Aquarius warm dark matter simulations. We also thank Tameem Adel for discussions at an early stage of this project. We acknowledge the support of the D-ITP consortium, a programme of the Netherlands Organization for Scientific Research (NWO) that is funded by the Dutch Ministry of Education, Culture and Science (OCW). N.~Bozorgnia is grateful to the Institute for Research in Fundamental Sciences in Tehran for their hospitality during her visit. G.B. (PI) and N.~Bozorgnia acknowledge support from the European Research Council through the ERC starting grant WIMPs Kairos.  J.~Bovy acknowledges the support of the Natural Sciences and Engineering Research Council of Canada (NSERC), funding reference number RGPIN-2015-05235, and from an Alfred P. Sloan Fellowship. N.~Bozorgnia  has received support from the European Union's Horizon 2020 research and innovation programme under the Marie Sklodowska-Curie grant agreement No 690575.

\begin{appendices}

\section{Scale radius of warm dark matter subhalos}
\label{sec:scaleradius}

In this appendix we discuss the scale radius of WDM subhalos and how it compares to that of CDM subhalos. As it was explained in section~\ref{subsec:subhaloimpacts}, we use eq.~\eqref{eq:rs} for the scale radius of CDM subhalos, which is obtained assuming Hernquist density profiles for the subhalos in the Via Lactea II catalog.

In order to compare the properties of WDM subhalos to their CDM counterparts, we use the density profiles of subhalos available from the Virgo consortium Aquarius project~\cite{Lovell2013} where in addition to a CDM simulation, four haloes with WDM particle mass of 1.5, 1.6, 2.0, and 2.3 keV are simulated. These include subhalos which are not spurious (see ref.~\cite{Lovell2013} for a detailed discussion of removing spurious subhalos from their halo catalogs) and are within 2 Mpc of the host halo center. We fit a Hernquist profile to the density profiles of CDM and WDM subhalos and check how the best fit scale radii change as a function of subhalo mass. 

To find the best fit Hernquist scale radius, for each individual simulated subhalo, we minimize the following $\chi^2$ function:
\begin{equation}
\chi^2(r_s) = \sum_{i}^{N} \frac{(\rho_i - \rho_{\rm Hern}(r_i, r_s))^2}{\sigma_i^2},
\end{equation}
where $\rho_i$ is the value of the DM density of the simulated subhalo at the radial bin $i$ with $r_i$ denoting the bin center, $\sigma_i$ is the corresponding $1\sigma$ Poisson error, $N$ is the number of radial bins we consider to perform the fit, and $\rho_{\rm Hern}(r_i, r_s)$ is the Hernquist density profile evaluated at radius $r_i$ as a function of the scale radius $r_s$. 

The density profiles are computed in spherical shells spaced equally in $\log(r)$. We find the inner and outer radii for performing the fit by using the criteria discussed in Springel {\it et al.}~\cite{Springel2008}. Namely, the inner radius for the fit, $r_{\rm min}$, is set to the radius in which convergence is achieved according to the Power {\it et al.} criterion~\cite{Power:2002sw}. The outer radius, $r_{\rm max}$, is set to the largest radius where the density of bound mass is more than 80\% of the total mass density. 

After specifying the radial range for the fit, we introduce the following two criteria to find \emph{good} subhalos for performing the fit: {\it (i)} $r_{\rm max} - r_{\rm min} > 1$~kpc, and {\it (ii)} the number of bins is greater or equal to 5. These additional criteria ensure that we are not fitting over a small radial range and we have enough bins for performing the fit. Finally, we need to consider the effect of tidal stripping on the goodness of fit. With the criteria mentioned, we don't obtain a good fit for some subhalos since we are probing a radial range where tidal stripping becomes important and there is a sharp decrease in the DM density from one bin to the next. To avoid this, we set the last bin we consider, $i_f$, as the bin where the DM density in the next bin decreases by more than an order of magnitude, i.e.~where $\rho_{i_f+1} < 0.1~\rho_{i_f}$. With these criteria we retain 4482 CDM subhalos over the mass range of $[1.5 \times 10^6, 10^{11}]~\Msun$. Notice that the lower boundary of the subhalo mass range is set by the requirement that the subhalo has at least 100 particles. 

\begin{figure}[t]
\centering
\includegraphics[scale=0.7]{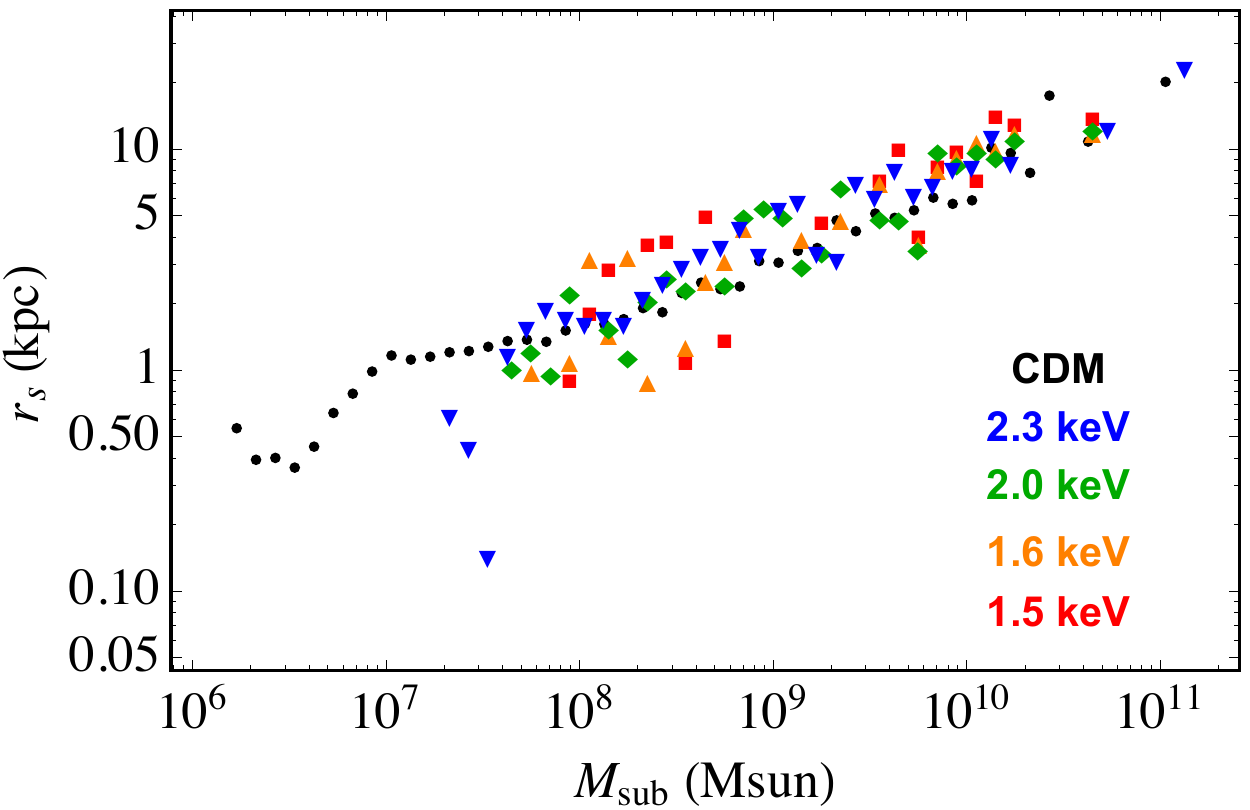}
\caption{The mean values of the best fit Hernquist scale radii of subhalos in the CDM and WDM Aquarius simulations as a function of the subhalo mass. The halos in the WDM simulations have WDM particle masses in the range 1.5 -- 2.3 keV.}
\label{fig:WDMrs}
\end{figure}

The result is shown in figure~\ref{fig:WDMrs} where we have plotted the mean values of the best fit Hernquist scale radius at different masses, considering equal size bins in $\log_{10}(M_{\rm sub})$. As it can be seen in the figure, the overall deviation between the best fit scale radii of CDM halos and WDM halos with different WDM particle mass is small. Although for $M_{\rm sub} \gtrsim 10^{9}~\Msun$, the scale radii for WDM halos are in general larger than those of CDM halos, for $M_{\rm sub} \leq 10^{9}~\Msun$ which is the subhalo mass range probed in our study there is no clear trend. Moreover, because the impact parameters considered in this work are much larger than the scale radius of the interacting subhalos with the streams, we use the same scale radius relation to describe both WDM and CDM subhalos. 

To capture the overall variation of the scale radius with the subhalo mass, we find the best fit $r_s(M_{\rm sub})$ relation from the mean scale radii of CDM subhalos in the Aquarius simulations shown in figure \ref{fig:WDMrs}, 
\begin{equation}
r_{\rm{s}}|_{\rm{fit}} = 1.24 ~\rm{kpc} \left(\frac{M_{\rm{sub}}}{10^8 \Msun}\right)^{0.39}.
\label{eq:rsfit}
\end{equation}

\begin{figure}[t]
\centering
\includegraphics[scale=0.55]{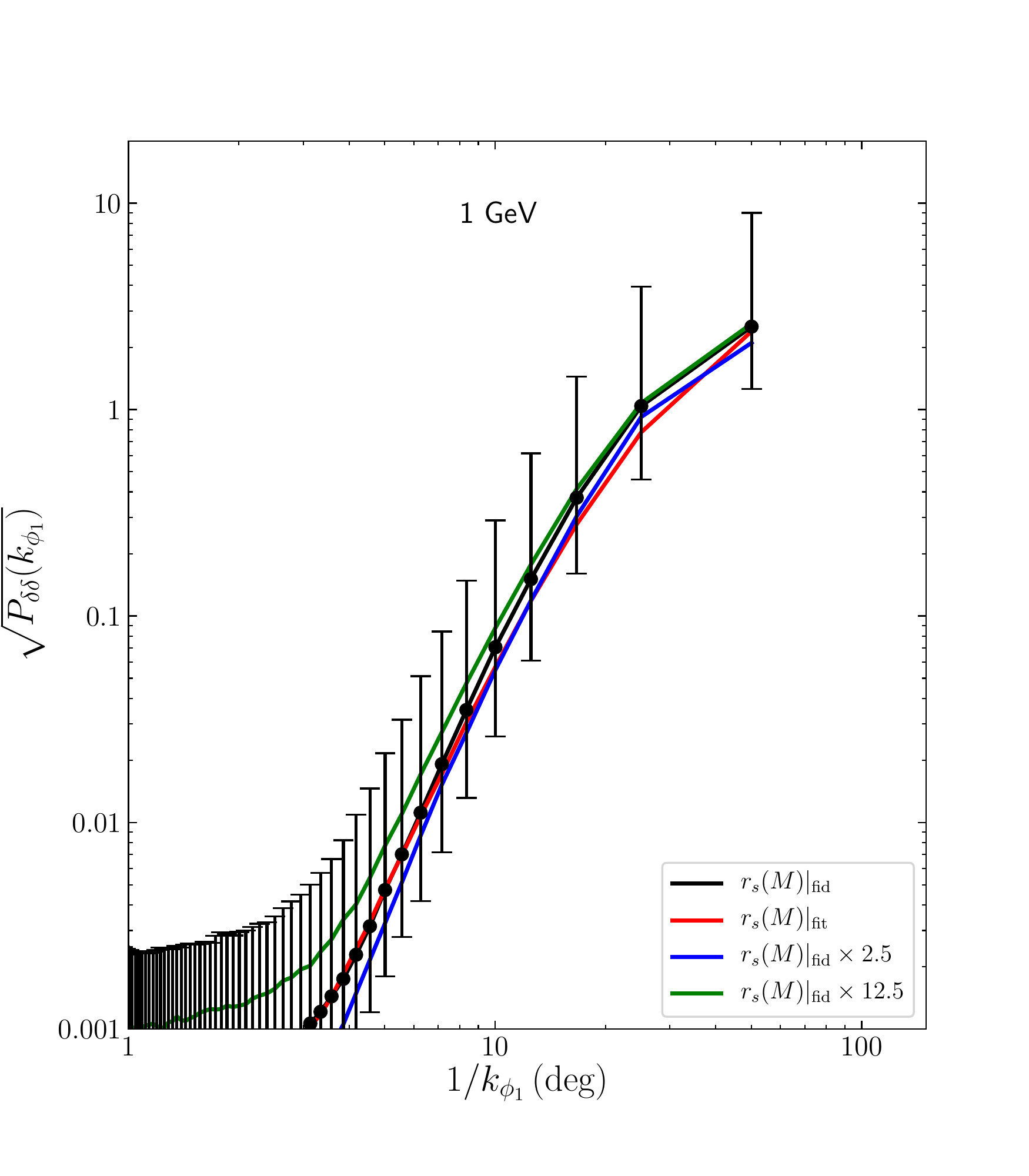}
\caption{Comparison of the median density power spectrum in the 1 GeV CDM scenario with different choices for the scale radius. The black curve is for the fiducial scale radius relation given by eq.~\eqref{eq:rs}, the red curve is with the scale radius relation found in eq.~\eqref{eq:rsfit} from a fit to the scale radii of subalos in the Aquarius simulations. In both of these cases the maximum impact parameter $b_{\rm{max}}$ is fixed to 5 times the scale radius for a particular mass subhalo. The error bars represent the $1\sigma$ dispersion of the power spectrum around the median for the fiducial case. The blue and green curve shows the cases when the fiducial scale radii of the subhalos are increased and decreased 2.5 times respectively while keeping the maximum impact parameter the same as the fiducial case. } 

\label{fig:CDMrs_Pk}
\end{figure}

This equation is slightly different from eq.~\eqref{eq:rs} which is our fiducial scale radius relation. In figure \ref{fig:CDMrs_Pk} we show how this difference in the scale radius relation will affect our analysis. The black solid line is the median power of 2500 simulations with the fiducial scale radius relation given by eq.~\eqref{eq:rs}. The error bars represent the $1\sigma$ dispersion of these simulations. The red curve is the median power of the simulations with the new scale radius fit given by eq.~\eqref{eq:rsfit}. The green and blue curves are the median power spectrum with 0.4 and 2.5 times scale radius but keeping the maximum impact parameter $b_{\rm{max}}$ fixed for a particular mass subhalo. All the power spectra are within $1\sigma$ dispersion of the fiducial case, suggesting that varying the scale radius within what we discussed will not affect our analysis.

\end{appendices}

\FloatBarrier

\bibliographystyle{JHEP}
\bibliography{refs}

\end{document}